\documentclass[onecolumn]{aa} 

\usepackage{graphicx} 
\usepackage{epsfig}

\newcommand{\figeq}[3]{\raisebox{#1}{\includegraphics[width=#2]{#3.eps}}}
\newcommand{\pl}{\partial} 
\renewcommand{\d}{{\rm d}} 
 
\newcommand{\beq}{\begin{equation}} 
\newcommand{\eeq}{\end{equation}} 
\newcommand{\beqa}{\begin{eqnarray}} 
\newcommand{\eeqa}{\end{eqnarray}} 
\newcommand{\bea}{\begin{array}} 
\newcommand{\ea}{\end{array}} 
\newcommand{\cG}{{\cal G}} 
\newcommand{\rhob}{\overline{\rho}} 
\newcommand{\lag}{\langle} 
\newcommand{\rag}{\rangle}
\newcommand{\bx}{{\bf x}}
\newcommand{\bp}{{\bf p}}
\newcommand{\df}{\delta \! f}
\newcommand{\bk}{{\bf k}}
\newcommand{\bw}{{\bf w}}

\newcommand{\cO}{{\cal O}}
\newcommand{\tK}{{\tilde K}}
\newcommand{\dLo}{\delta_{L0}}
\newcommand{\dL}{\delta_L}
\newcommand{\PLo}{P_{L0}}
\newcommand{\bpL}{{\bf p}_L}

\newcommand{\DL}{\Delta_L}
\newcommand{\Det}{\mbox{Det}}
\newcommand{\Tr}{\mbox{Tr}}
\newcommand{\tKs}{{\tilde K}_{\rm s}}
\newcommand{\bF}{{\bf F}}
\newcommand{\ti}{t_{\rm i}}
\newcommand{\etai}{\eta_{\rm i}}

\newcommand{\fb}{\overline{f}}
\newcommand{\Ks}{K_{\rm s}}

\newcommand{\Di}{\Delta_{\rm i}}
\newcommand{\Dim}{\Delta_{\rm i}^{-1}}
\newcommand{\tGam}{{\tilde \Gamma}}
\newcommand{\phis}{\phi_{\rm s}}
\newcommand{\Go}{G_0}
\newcommand{\Gom}{G_0^{-1}}
\newcommand{\phib}{\overline{\phi}}
\newcommand{\Gm}{G^{-1}}
\newcommand{\Dm}{\Delta^{-1}}
\newcommand{\tS}{{\tilde S}}
\newcommand{\Sint}{S_{\rm int}}
\newcommand{\lambdab}{\overline{\lambda}}
\newcommand{\cGm}{{\cal G}^{-1}}
\newcommand{\self}{\Phi}
\newcommand{\tR}{R^T}
\newcommand{\tG}{{\tilde G}}
\newcommand{\tSigma}{{\Sigma}^T}
\newcommand{\tGamtwo}{{\tilde \Gamma}_{\rm 2 loops}}
\newcommand{\Sigmatwo}{\Sigma_{\rm 2 loops}}
\newcommand{\Pitwo}{\Pi_{\rm 2 loops}}

\newcommand{\Db}{\overline{\Delta}}
\newcommand{\Dbm}{\overline{\Delta}^{-1}}
\newcommand{\cN}{{\cal N}} 
\newcommand{\ux}{\underline{x}} 
\newcommand{\etaib}{\underline{\eta}_{\rm i}}

\newcommand{\etabi}{\overline{\eta}_{\rm i}}

\begin{document} 
 
\title{A new approach to gravitational clustering: a path-integral formalism and large-$N$ expansions}    
\author{P. Valageas}   
\institute{Service de Physique Th\'eorique, CEN Saclay, 91191 Gif-sur-Yvette, 
France}  
\date{Received 20 June 2003/ Accepted 4 March 2004} 
 
\abstract{ 
We show that the formation of large-scale structures through gravitational instability in the expanding universe can be fully described through a path-integral formalism. We derive the action $S[f]$ which gives the statistical weight associated with any phase-space distribution function $f(\bx,\bp,t)$. This action $S$ describes both the average over the Gaussian initial conditions and the Vlasov-Poisson dynamics. Next, applying a standard method borrowed from field theory we generalize our problem to an $N-$field system and we look for an expansion over powers of $1/N$. We describe three such methods and we derive the corresponding equations of motion at the lowest non-trivial order for the case of gravitational clustering. This yields a set of non-linear equations for the mean $\fb$ and the two-point correlation $G$ of the phase-space distribution $f$, as well as for the response function $R$. These systematic schemes match the usual perturbative expansion on quasi-linear scales but should also be able to treat the non-linear regime. Our approach can also be extended to non-Gaussian initial conditions and may serve as a basis for other tools borrowed from field theory.
 } 
 
\maketitle

\section{Introduction} 
\label{Introduction}

In standard cosmological scenarios large-scale structures in the universe have formed through the growth of small initial density fluctuations by gravitational instability, see \cite{Peeb1}. Once they enter the non-linear regime, high-density fluctuations build massive objects which will give birth to galaxies or clusters. Thus, it is important to understand the dynamics of these density perturbations in order to describe the large-scale structures we observe in the present universe. Moreover, observations (e.g. weak-lensing or galaxy surveys) can provide a measure of the power-spectrum of the density fluctuations down to non-linear scales. Hence it is useful to devise theoretical tools which may treat the non-linear regime to make the connection between the data and the primordial power-spectrum.

However, the non-linear Vlasov-Poisson equations that describe the dynamics of such density fluctuations within an expanding universe are not easy to solve and most rigorous approaches are based on perturbative expansions. More precisely, the distribution of matter is usually described as a pressure-less fluid through the standard equations of hydrodynamics (continuity and Euler equations) coupled to the Poisson equation for the gravitational potential. Then, one often looks for a solution of these equations of motion in the form of a perturbative expansion over powers of the initial fluctuations, or rather over powers of the linear growing mode (e.g., \cite{Fry1}, \cite{Gor1}, \cite{Ber2}). An alternative method which is not based on such perturbative expansions nor on the hydrodynamical approximation, developed in \cite{Val2}, can go somewhat beyond these previous approaches for the statistics of the density within spherical cells. However, all these methods break down at shell-crossing and cannot handle the non-linear regime. Hence they are restricted to large scales, except for the approach described in \cite{Val4} which can still handle very rare voids in the non-linear regime (where shell-crossing is not important).

An alternative approach to this problem is to investigate simplified equations of motion which would capture most of the physics and would be easier to analyze. Thus, the well-known Zel'dovich approximation (\cite{Zel}) extrapolates the trajectories of the particles from the linear regime. Next, the density field is reconstructed from the particles displacements. The advantage of looking at the trajectories is that one can handle large density perturbations and even go beyond shell-crossing: particles may cross each other and their trajectories remain well-defined while the local density momentarily diverges (which prevents the application of naive perturbative methods to the density field itself). Numerical studies have shown that the Zel'dovich approximation works well until shell-crossing and it can reproduce the large-scale skeleton of the matter distribution (e.g., \cite{Coles1}). However, it breaks down after shell-crossing since it does not include any feedback: particles keep moving ahead after pancake formation and are not pulled back by the gravitational potential well. This has led to the adhesion model (\cite{Gur1}, \cite{Ver1}) where one adds an artificial viscosity term to the Euler equation so as to mimic the gravitational ``sticking'' of particles within potential wells. Then, the thickness of filaments or halos depends on this viscosity parameter. This model can be generalized by including a dependence on density and time within the viscosity term (\cite{Buch1}) which may improve the predictions obtained for the thickness of non-linear objects.  However, it remains to be seen whether one can describe the inner structure of non-linear halos by following this line of attack.

A different approach was proposed by \cite{Wid1} who suggested to replace the Vlasov-Poisson system by a Schr$\ddot{\rm o}$dinger-Poisson system. This model is much closer to the original Vlasov-Poisson system than the hydrodynamical approaches recalled above and it contains more phase-space information. Indeed, from the wave-function $\psi(\bx,t)$ one may construct a full distribution function $f(\bx,\bp,t)$ (which is not restricted to a Dirac or a Maxwellian in velocity space) which follows an equation of motion that reduces to the Vlasov equation in some limit (see \cite{Wid1}). The advantage of using $\psi(\bx,t)$ is that it reduces the phase space from six coordinates $(\bx,\bp)$ down to three $(\bx)$. As described in \cite{Wid1} and \cite{Dav1}, this approach can treat shell-crossing and virialization. A comparison with the Zel'dovich approximation and simple examples show that this method is indeed a promising tool for cosmological purposes (\cite{Coles2}).

In this article, we develop a new approach to this problem by working directly with the original Vlasov-Poisson equations. For cosmological purposes, we are not really interested in the evolution of the system for a few specific initial conditions but we rather wish to derive the statistical properties of the system after averaging over these initial conditions (usually taken to be Gaussian from inflationary scenarios). This suggests the use of a functional approach as in usual statistical physics, which directly focuses on the correlation functions of the system (or other statistics). However, in order to obtain such quantities one must resort to approximation schemes since usually non-Gaussian path-integrals cannot be computed exactly. Not to mix up the sources of error introduced by such approximations (associated with the procedure used to derive averaged quantities) with those entailed by the use of approximate equations of motion it is desirable to first apply this approach to the exact Vlasov-Poisson system. Moreover, the ultimate goal is to be able to describe the original system itself. However, our study is also complimentary to previous works which investigated hydrodynamical or ``quantum'' systems as a substitute to the Vlasov-Poisson equations since it could be applied to such approaches.

Thus, we show that the statistical properties of the system, assuming Gaussian initial conditions, are fully described by a path-integral formalism and we derive the action $S[f]$ which gives the statistical weight $e^{-S[f]}$ associated with any phase-space distribution function $f(\bx,\bp,t)$. This action $S$ contains both the average over the Gaussian initial conditions and the Vlasov-Poisson dynamics. Moreover, we note that it yields the response functions of the system in addition to the correlation functions. Then, this result allows us to use standard tools of field theory. In particular, in this article we describe three such methods, based on ``large-$N$'' expansions, and we derive for the case of gravitational clustering the equations of motion they imply for the mean phase-space distribution $\fb$ and its two-point correlation $G$, at next-to-leading order (the first non-trivial order).

This paper is organized as follows. In Sect.\ref{Equations of motion} we recall the Vlasov-Poisson equations of motion and we introduce our notations. We also pay some attention to the infrared divergences which appear for power-law linear power-spectra $P_L(k) \propto k^n$ with $n \leq -1$. Next, in Sect.\ref{Functional formulation} we show how our system can be described through a path-integral formalism and we derive the corresponding action $S[f]$. We also discuss the response functions of the system and we briefly review several approximation schemes. Then, in Sect.\ref{Direct steepest-descent method} we derive up to next-to-leading order the equations given by a direct steepest-descent method. We discuss in Sect.\ref{1PI Effective action} an alternative 1PI effective action scheme and we finally obtain in Sect.\ref{2PI effective action} the Schwinger-Dyson equations given by a third approach based on the 2PI effective action (we also present in App.\ref{Reduced 2PI effective action} a simple proof of a key result needed in this approach). We also show that the infrared behaviour encountered in the case of large-scale structure formation selects some approaches over others. Finally, we investigate a standard perturbative expansion in Sect.\ref{Perturbative expansion} and we discuss our results in Sect.\ref{Conclusion}.

\section{Equations of motion}
\label{Equations of motion}

\subsection{Vlasov-Poisson system}
\label{Vlasov-Poisson system}

First, we recall here the equations of motion which
describe the gravitational dynamics of a collisionless fluid in an
expanding universe. We consider homogeneous initial conditions, that is the
system is statistically invariant through translations. In the continuous 
limit where the mass $m$ of the
particles goes to zero the system is fully described by the
distribution function (phase-space density) $f(\bx,\bp,t)$ where we
note $\bx$ the comoving coordinate of a particle and $\bp$ its momentum (which we shall also refer to as the velocity) defined by:
\beq
\bp = a^2 \dot{\bx} , \hspace{0.3cm} \mbox{and we normalize $f$ by: } \hspace{0.3cm} \rho(\bx,t) = \rhob \int f(\bx,\bp,t) \; \d\bp    .
\label{p1}
\eeq
Here we introduced the scale factor $a(t)$, the comoving density $\rho(\bx,t)$ and the mean comoving density of the universe $\rhob$ (which is independent of time). Note that we normalized the distribution function $f(\bx,\bp,t)$ by the constant mass density $\rhob$, so that $\int f \d\bp$ is dimensionless. Then, the gravitational dynamics is given by the collisionless Boltzmann equation, also called Vlasov equation, coupled with the Poisson equation (\cite{Peeb1}):
\beq
\frac{\pl f}{\pl t} + \frac{\bp}{a^2} . \frac{\pl f}{\pl \bx} -
\frac{\pl \Phi}{\pl \bx} . \frac{\pl f}{\pl \bp} = 0 , \hspace{0.4cm}
\mbox{with} \hspace{0.4cm} \Delta \Phi = \frac{4\pi \cG}{a} (\rho -
\rhob) = \frac{4\pi \cG}{a} \; \rhob \; \delta  \hspace{0.4cm} \mbox{and} \hspace{0.4cm} \delta(\bx,t) = \frac{\rho(\bx,t)-\rhob}{\rhob} ,
\label{Bol1}
\eeq
where $\Phi$ is the gravitational potential and $\delta$ is the density contrast. The fact that the Poisson equation involves the density perturbation $(\rho-\rhob)$ rather than the total density $\rho$ comes from the change from physical coordinates to comoving coordinates (this also removes the cosmological constant term $\Omega_{\Lambda}$ from these equations of motion). In order to derive eq.(\ref{Bol1}) we neglected the discrete character of the distribution of matter and we assumed small-scale smoothness (i.e. there is a small-scale cutoff to the power-spectrum of the density fluctuations), see \cite{Peeb1}. As can be seen from eq.(\ref{Bol1}), the gravitational force $\bF(\bx,t)$ can be written as:
\beq
\bF(\bx,t) = - \frac{\pl \Phi}{\pl \bx} = \frac{\cG}{a} \int \d \bx' \; (\rho(\bx')-\rhob) \; \frac{\bx'-\bx}{|\bx'-\bx|^3} = \frac{\cG}{a} \int \d \bx' \; \rhob \delta(\bx') \; \frac{\bx'-\bx}{|\bx'-\bx|^3} = \frac{\cG}{a} \int \d \bx' \; \rho(\bx') \; \frac{\bx'-\bx}{|\bx'-\bx|^3} .
\label{Fgrav1}
\eeq
The fourth term in eq.(\ref{Fgrav1}) merely uses the definition of the density contrast $\delta(\bx,t)$ while the fifth term uses the fact that the integration over the angles of $(\bx'-\bx)$ of the factor multiplied by the constant $\rhob$ vanishes (see \cite{Peeb1}). Note that this actually involves an implicit temporary regularization of the gravitational interaction at large scales (see eq.(\ref{kc1}) below). This is sometimes referred to as the ``Jeans swindle''. For practical purposes, it is convenient at some points to work in global Fourier space. Thus, we introduce the Fourier conjugates $\bk$ and $\bw$ of the spatial coordinate $\bx$ and of the momentum $\bp$, and we define the Fourier transforms of the distribution function $f(\bx,\bp,t)$ and of the density contrast $\delta(\bx,t)$ by:
\beq
f(\bx,\bp,t) = \int \d\bk\d\bw \; e^{i(\bk.\bx+\bw.\bp)} \; f(\bk,\bw,t) \hspace{0.4cm} \mbox{and} \hspace{0.4cm} \delta(\bx,t) = \int \d\bk \; e^{i\bk.\bx} \; \delta(\bk,t) .
\label{Four1}
\eeq
To avoid introducing too many variable names, we use the same notations
both in real space $(\bx,\bp)$ and in Fourier space $(\bk,\bw)$. The arguments
should remove any ambiguity. We also define the Fourier transforms of other
fields as in eq.(\ref{Four1}). In Fourier space, the gravitational force can be obtained from eq.(\ref{Bol1}) or eq.(\ref{Fgrav1}) and we write:
\beq
\bF(\bk,t) = i \frac{4\pi\cG}{a} \frac{\bk}{k^2} \rho(\bk,t) = i \frac{4\pi\cG\rhob}{a} (2\pi)^3 \frac{\bk}{k^2} f(\bk,0,t) .
\label{Fgrav2}
\eeq
Note that we could have chosen the density perturbation $(\rho-\rhob)$ rather than the total density $\rho$ in eq.(\ref{Fgrav2}), as can be seen from eq.(\ref{Fgrav1}). This merely translates the fact that the gravitational potential is not fully defined by the Poisson equation (one should add boundary conditions). This is directly related to the ``Jeans swindle'' discussed below eq.(\ref{Fgrav1}) which also leads to the ``property'':
\beq
\frac{\bk}{k^2} \; \delta_D(\bk) = 0 , \hspace{0.4cm} \mbox{because} \hspace{0.2cm} \frac{\bk}{k^2} \hspace{0.2cm} \mbox{stands for} \hspace{0.2cm} \lim_{k_c \rightarrow 0} \frac{\bk}{k^2 + k_c^2} ,
\label{kc1}
\eeq
where $\delta_D$ stands for Dirac's distribution and we wrote explicitly the large-scale regularization of the gravitational interaction (beyond the scale $1/k_c \rightarrow \infty$). In practice we do not need to write explicitly the cutoff $k_c$ and all our equations are written with $k_c=0$ (i.e. no large-scale cutoff) because we usually do not encounter ambiguous quantities as in (\ref{kc1}). The only exceptions are eq.(\ref{trace1}) and eq.(\ref{fb1}) below where we must remember the ``property'' (\ref{kc1}). Then, using eq.(\ref{Fgrav2}) the Vlasov equation (\ref{Bol1}) reads in Fourier space:
\beq
\frac{\pl f}{\pl t} - \frac{\bk}{a^2} . \frac{\pl f}{\pl \bw} = \frac{4\pi\cG\rhob}{a} (2\pi)^3 \int \d\bk' \; \frac{\bk'.\bw}{k'^2} \; f(\bk',0) f(\bk-\bk',\bw) .
\label{Bol2}
\eeq
This {\it non-linear} and {\it non-local} equation describes the
dynamics of the collisionless fluid. Thus, in order to investigate the
formation of large-scale structures in the universe (neglecting the
influence of baryons onto the dark matter) we could solve
eq.(\ref{Bol2}) for some specific initial conditions and next perform a
suitable average over these initial conditions (characterized by the
power-spectrum of the early linear perturbations). In the following, introducing the matrix $\cO$ and the vertex $K$, we shall write the quadratic equation (\ref{Bol2}) in the more concise form :
\beq
\cO(x_1,x_2) f(x_2) = K(x_1;x_2,x_3) f(x_2) f(x_3) + \etai(x_1) ,  \hspace{0.3cm} \mbox{with} \hspace{0.1cm} f(x_1) = \theta(t_1-\ti) f(x_1) \hspace{0.1cm} \mbox{and} \hspace{0.1cm} \etai(x_1) = \delta_D(t_1-\ti) \etaib(\ux_1) ,
\label{Bol3}
\eeq
where the argument $x=(\bk,\bw,t)=(\ux,t)$ (or $x=(\bx,\bp,t)$ in real space) represents the space, velocity and time variables, the set of all variables excluding time is denoted by underlining the argument (e.g., $\ux$); and we used the integration convention for repeated indices. The new term $\etai(x_1)$ in the r.h.s. in eq.(\ref{Bol3}) describes the initial conditions at time $\ti$ so that $f(x)=0$ for $t<\ti$ and $f(\ux,\ti)=\etaib(\ux)$. We shall eventually take the limit $\ti \rightarrow 0$. Throughout this article we note $\delta_D$ and $\theta$ the Dirac and Heaviside functions.

\subsection{Initial conditions}
\label{Initial conditions}

In this paper we consider standard cosmological scenarios where large-scale structures have formed through the growth of small density fluctuations by gravitational instability. Therefore, at early times the system shows small perturbations from the homogeneous expanding universe and one can use linear theory. In practice, this regime is usually studied within the framework of the hydrodynamical approximation where all particles at a given point $\bx$ have the same momentum $\bp(\bx)$ and the linear growing mode is given in partial Fourier space by (\cite{Peeb1}):
\beq
\dL(\bk,t) = D_+(t) \dLo(\bk)  \hspace{0.2cm} \mbox{and} \hspace{0.2cm}  \bpL(\bk,t) = i a^2 \dot{D}_+ \; \frac{\bk}{k^2} \; \dLo(\bk) , \hspace{0.2cm} \mbox{with} \hspace{0.2cm} \ddot{D}_+ + 2 \frac{\dot{a}}{a} \dot{D}_+ = \frac{4\pi\cG\rhob}{a^3} \; D_+ .
\label{dL1}
\eeq
Here $\dL$ and $\bpL$ are the linear density contrast and momentum while $\dLo(\bx)$ is the linear density contrast today (at $z=0$) and $D_+(t)$ is the linear growth factor. Next, one can look for a solution to the Euler-Poisson system through a perturbative expansion over powers of the linear growing mode $\dL$ (e.g., \cite{Ber1} for a review). As seen in \cite{Val1}, the same procedure can be applied to the Vlasov-Poisson system and it yields identical results at all orders (which strongly suggests that shell-crossing is beyond the reach of such perturbative schemes). The linear growing mode $\eta_L$ of the Vlasov equation which corresponds to (\ref{dL1}) is given by (see \cite{Val1}):
\beq
\eta_L(\bx,\bp,t) = \delta_D(\bp) + \dL(\bx,t) \delta_D(\bp) -  \bpL(x,t).\frac{\pl \delta_D}{\pl \bp}(\bp) \hspace{0.2cm} \mbox{and} \hspace{0.2cm} \eta_L(\bk,\bw,t) = \frac{\delta_D(\bk)}{(2\pi)^3} + \frac{D_+}{(2\pi)^3} \; \dLo(\bk) \left[ 1 + \frac{\dot{D}_+ a^2}{D_+} \; \frac{\bk.\bw}{k^2} \right] 
\label{etaL1}
\eeq
Here we included the constant term $\delta_D(\bp)$ which corresponds to the homogeneous expanding universe. Thus, at early times (or large scales) we have $f \rightarrow \eta_L$. The Dirac factors $\delta_D(\bp)$ in eq.(\ref{etaL1}) translate the fact that we have no velocity dispersion at linear order (we have a pressure-less hydrodynamical fluid) and higher orders of the perturbative expansion over powers of the linear mode $\eta_L$ merely generate higher-order derivatives of $\delta_D(\bp)$ (see \cite{Val1}). Then, we can choose for the initial conditions $\etai$ introduced in eq.(\ref{Bol3}):
\beq
\etaib(\bx,\bp) = \eta_L(\bx,\bp,\ti) , \hspace{0.2cm} \mbox{and we note} \hspace{0.2cm} \etabi = \lag \etai \rag  \hspace{0.2cm} \mbox{and} \hspace{0.2cm}  \Di(x_1,x_2) = \lag \etai(x_1)\etai(x_2) \rag_c ,
\label{Di1}
\eeq
where $\lag .. \rag$ denotes the average over the initial conditions. In this article we shall consider Gaussian initial conditions, so that both the linear density field $\dL$ and the linear growing mode $\eta_L$ are Gaussian:
\beq
\lag \dLo(\bk_1) \dLo(\bk_2) \rag = \PLo(k_1) \delta_D(\bk_1+\bk_2) \hspace{0.2cm} \mbox{and we note} \hspace{0.2cm} \DL(x_1,x_2) = \lag \eta_L(x_1)\eta_L(x_2) \rag_c .
\label{PLo1}
\eeq
Then, the field $\etai$ is also Gaussian and its connected two-point correlation $\Di$ can be expressed in terms of the linear power-spectrum $\PLo(k)$ through eq.(\ref{etaL1}). Note that for a non-zero initial time $\ti >0$ the choice (\ref{Di1}) for the initial conditions $\etaib$ actually yields both growing and decaying linear modes. However, in the limit $\ti \rightarrow 0$ we indeed recover the pure linear growing mode $\eta_L$ defined in eq.(\ref{etaL1}).

\subsection{Long-wavelength divergences}
\label{Long-wavelength divergences}

From eq.(\ref{dL1}), one can see that for a power-law linear power-spectrum $\PLo(k) \propto k^n$ (with $-3<n<1$ in the cosmological context) the rms linear momentum at scale $R$ formally scales as $\sigma_p(R) \propto R^{-(n+1)/2}$. This actually leads to infrared divergences for the linear velocity field $\bpL$ as defined in eq.(\ref{dL1}) for $n\leq-1$. More precisely, if we try to set up some generic initial conditions from eq.(\ref{dL1}) we can see that long wavelengths yield a divergent contribution to $\bpL(\bx)$. Thus, we can no longer use eq.(\ref{dL1}) to define the linear velocity field. Within the standard perturbative approach based on the Euler-Poisson system, these infrared divergences actually cancel out (at least at leading order) when one only considers the equal-time correlation of the density field, as seen in \cite{Jain1} (also \cite{Vish1}). As pointed out by these authors, this can be understood from the fact that contributions to the velocity field from long-wave modes correspond to an almost uniform translation of the fluid and therefore should not affect the growth of density structures on small scales. However, these infrared contributions can no longer be ignored when one studies different-times correlations or the velocity field itself, which is clearly the case when we work with the Vlasov equation.

As suggested in \cite{Jain1}, the only quantity which really matters is the velocity difference between neighbouring points and not the overall mean velocity. This is not surprising since the equations of motion written in real physical space obey the usual Galilean symmetry (note that this is no longer the case for eq.(\ref{Bol1}) or eq.(\ref{Bol2}) which are written in comoving coordinates). Therefore, we must recover the same physical behaviour if we subtract the initial velocity field by some constant. In particular, we could normalize the linear velocity so that $\bpL(\bx=0)=0$ for instance. Thus, we now define the linear velocity field $\bpL(\bx)$ by:
\beq
\bpL(\bx,t) = \int \d\bk \; \left(e^{i\bk.\bx}-1\right) i a^2 \dot{D}_+ \; \frac{\bk}{k^2} \; \dLo(\bk) .
\label{pL1}
\eeq
One can see that this still yields a linear solution to the Vlasov equation provided we subtract a constant term over space to the gravitational force so that $\bF(\bx=0)=0$. Of course, this simple change of normalization does not modify the physics of gravitational clustering and it is consistent with Poisson's equation. Finally, this yields for the Vlasov equation:
\beq
\frac{\pl f}{\pl t} - \frac{\bk}{a^2} . \frac{\pl f}{\pl \bw} = \frac{4\pi\cG\rhob}{a} (2\pi)^3 \int \d\bk' \; \frac{\bk'.\bw}{k'^2} \; f(\bk',0) \left[ f(\bk-\bk',\bw) - f(\bk,\bw) \right] .
\label{Bol6}
\eeq
By comparison with eq.(\ref{Bol2}) we see that the ``counter-term'' $f(\bk,\bw)$ in the bracket ensures that the contribution from large scales (i.e. $k' \rightarrow 0$) converges (when we work in real space $\bx$). Therefore, eq.(\ref{Bol6}) allows us to investigate linear power-spectra with $n \leq -1$. Let us stress that eq.(\ref{Bol6}) is exact and is strictly equivalent to eq.(\ref{Bol2}) with respect to gravitational clustering, that is the density fields obtained from both evolution equations are identical. However, the velocities and the positions of the particles differ between both frameworks by some values which are constant over space (but change with time). These offsets actually diverge for $n \leq -1$ because of long wavelength contributions but they have no influence on the physics at work. Working with eq.(\ref{Bol6}) ensures that when we investigate the physics at some scale $R$ we only encounter quantities which are governed by this scale (or possibly the scale which is turning non-linear), and not by some much larger scale (where $n$ crosses the threshold $-1$) which eventually disappears as various contributions cancel out. This is a priori of great interest for numerical purposes. Unfortunately, within this framework we lost homogeneity (i.e. invariance through translations) since we singled out the origin $\bx=0$. This makes the analysis of the problem and the interpretation itself of the correlations of the density field more intricate. Therefore, in the following we shall only consider the case of homogeneous initial conditions (i.e. with a large-scale cutoff) which should be sufficient for practical purposes. However, the influence of these infrared divergences will play a key role in Sect.\ref{Infrared divergences} as they will discriminate between various approximation schemes.

\section{Functional formulation}
\label{Functional formulation}

\subsection{Derivation of the actions $S[f]$ and $S[f,\lambda]$, mean response functions}
\label{Derivation of the actions}

In practice, we are not really interested in the exact solution $f$ of the equation of motion for a given initial condition $\etai$. Indeed, since the initial condition $\etai$ is a random field we are mainly interested in the statistical properties of the phase-space distribution function $f$. They can be described through a Heisenberg operator formalism (\cite{MSR}) or a functional approach (e.g., \cite{Phy}, \cite{Jens}). We shall apply to the Vlasov-Poisson equation (\ref{Bol2})-(\ref{Bol3}) the path-integral formalism which is more convenient. Hereafter we work in real space $(\bx,\bp,t)$ (except in eqs.(\ref{Bolint1})-(\ref{tK1}) below and in Sect.\ref{Perturbative expansion}). Thus, let us consider any functional $F(f)$ of the classical field $f$ (e.g., an $n-$point correlation). Its average over the initial conditions can be computed as:
\beq
\lag F(f) \rag = \int [\d \etai] \; F(f) \; e^{- \frac{1}{2} (\etai-\etabi) . \Dim . (\etai-\etabi)} = \int [\d f] \; | \Det M | \; F(f) \; e^{- \frac{1}{2} (\cO f - \Ks f^2 - \etabi) . \Dim . (\cO f - \Ks f^2 - \etabi)} ,
\label{F1}
\eeq
where we have taken the Gaussian average over $\etai$ and in the last term we made the change of variable $\etai \rightarrow f$, using eq.(\ref{Bol3}). We also introduced the symmetric vertex (over its last two arguments) $\Ks(x;x_1,x_2)=[K(x;x_1,x_2)+K(x;x_2,x_1)]/2$. We do not write explicitly the normalization factors which appear in front of path-integrals like (\ref{F1}). If needed, they can be recovered from the constraint $\lag 1 \rag=1$. The Jacobian $| \Det M |$, with $M=\delta \etai/\delta f$, is given by:
\beq
\Det M = \Det ( \cO - 2\Ks f ) = \Det(\cO) \Det(1-2\tKs f) = \Det(\cO) \exp [\Tr \ln (1-2\tKs f)] ,
\label{Det}
\eeq
where $1-2\tKs f$ is the functional derivative with respect to $f$ of the integral form of the Vlasov equation (\ref{Bol2})-(\ref{Bol3}) (see also \cite{Zinn1}). The latter can be integrated by the method of characteristics which yields:
\beqa
f(\bk,\bw,t) & = & \etaib\left( \bk,\bw+\bk\int_{\ti}^t \frac{\d t'}{a'^2} \right) \nonumber \\ & & + \; 4\pi\cG\rhob (2\pi)^3 \int_{\ti}^t \frac{\d t'}{a'} \int \d\bk' \; \frac{\bk'}{k'^2} . \left[ \bw+\bk\int_{t'}^t \frac{\d t''}{a''^2} \right] f(\bk',0,t') f\left( \bk-\bk',\bw+\bk\int_{t'}^t \frac{\d t''}{a''^2} , t' \right) ,
\label{Bolint1}
\eeqa
so that the vertex $\tK$ reads in Fourier space (with all times larger than $\ti$):
\beq
\tK(x;x_1,x_2) = \frac{4\pi\cG\rhob}{a(t_1)} (2\pi)^3 \theta(t-t_1) \;
\delta_D(t_1-t_2) \; \delta_D(\bw_1) \; \delta_D\left(\bw_2-\bw-\bk\int_{t_1}^{t} \frac{\d t'}{a'^2} \right) \; \frac{\bk_1.\bw_2}{k_1^2}  \; \delta_D(\bk_1+\bk_2-\bk) . 
\label{tK1}
\eeq
Of course, we have $\cO.\tK=K$, and we defined the symmetric vertex $\tKs(x;x_1,x_2)=[\tK(x;x_1,x_2)+\tK(x;x_2,x_1)]/2$. Next, we expand the logarithm in eq.(\ref{Det}) which yields:
\beq
\Tr \ln (1-2\tKs f) = - \sum_{q=1}^{\infty} \frac{1}{q} \; \Tr \; (2\tKs f)^q = - \Tr \; 2\tKs f = 0  \hspace{0.3cm} \mbox{since} \hspace{0.3cm} \int \d x \; \tKs(x;x,y) = 0 .
\label{trace1}
\eeq
Here we used the fact that $\Tr (2\tKs f)^q=0$ for $q \geq 2$ (see also \cite{Val1}, \cite{Zinn1}) because of causality (i.e. the Heaviside factor which appears in the kernel $\tKs$) and the last equality uses the property (\ref{kc1}) to remove any ambiguity. Therefore, the Jacobian in eq.(\ref{F1}) is equal to unity, up to the irrelevant constant $\Det(\cO)$. Thus, we obtain:
\beq
\lag F(f) \rag = \int [\d f] \; F(f) \; e^{-S[f]} \hspace{0.3cm} \mbox{with} \hspace{0.3cm} S[f] = \frac{1}{2} \left(\cO f-\Ks f^2 -\etabi\right) . \Dim . \left(\cO f-\Ks f^2 -\etabi\right) .
\label{Sf}
\eeq
Thus, eq.(\ref{Sf}) shows that our system is fully described by the action $S[f]$ therein defined. Note that $S[f]$ depends on the configuration of the field $f$ over phase-space $(\bx,\bp)$ and time $t$. In particular, it is clear that our problem is non-stationary: we study an out-of-equilibrium dynamics driven by gravitational instability. Note that the procedure described above could easily be extended to non-Gaussian initial conditions, provided the statistical weight of the field $\etai$ is still given by an action $S_{\etai}[\etai]$, which would no longer be quadratic. From eqs.(\ref{etaL1})-(\ref{PLo1}) one can easily show that the matrix $\Di$ is positive. Therefore, its inverse $\Dim$ is also positive, hence the path-integral (\ref{Sf}) is well-defined since $S[f]\geq 0$ (the action $S$ admits a lower bound).

As we shall see below, it is convenient to write the path-integral (\ref{Sf}) in the form:
\beq
\lag F(f) \rag = \int [\d f] [\d\lambda] \; F(f) \; e^{-S[f,\lambda]} \hspace{0.3cm} \mbox{with} \hspace{0.3cm} S[f,\lambda] = \lambda . ( \cO f-\Ks f^2 -\etabi ) - \frac{1}{2} \lambda . \Di . \lambda ,
\label{Sfl}
\eeq
where we introduced the auxiliary imaginary field $\lambda$ (i.e. $\lambda(x)= i \lambda'(x)$ with $\lambda'$ real). Indeed, we can check that the Gaussian integration over $\lambda$ of eq.(\ref{Sfl}) yields back eq.(\ref{Sf}). One interest of the formulation (\ref{Sfl}) is that the auxiliary field $\lambda$ generates the response functions of the system (\cite{Phy}). Indeed, let us add a small non-random perturbation $\zeta(x_1)$ to the right-hand side of the Vlasov equation (\ref{Bol2})-(\ref{Bol3}). This amounts to the change $\etai \rightarrow \etai+\zeta$ which can be described by the simple perturbation $\etabi \rightarrow \etabi+\zeta$. Therefore, in eq.(\ref{Sfl}) the action is changed to $S[f,\lambda] \rightarrow S[f,\lambda] - \lambda.\zeta$. Then, the functional derivatives with respect to $\zeta$ at the point $\zeta=0$ write:
\beq
\lag \frac{\delta F(f)}{\delta\zeta(x)} \rag_{\zeta=0} = \int [\d f] [\d\lambda] \; \lambda(x) \; F(f) \; e^{-S[f,\lambda]} , \hspace{0.3cm} \lag \frac{\delta^2 F(f)}{\delta\zeta(x_1)\delta\zeta(x_2)} \rag_{\zeta=0} = \int [\d f] [\d\lambda] \; \lambda(x_1) \lambda(x_2) \; F(f) \; e^{-S[f,\lambda]} , ..
\label{resp1}
\eeq
This clearly shows that insertions of the auxiliary field $\lambda$ yield the mean response functions of the system. In particular, the mean first-order response function $R(x_1,x_2)$ is:
\beq
R(x_1,x_2) = \left. \frac{\delta \lag f(x_1) \rag}{\delta\zeta(x_2)} \right|_{\zeta=0} = \lag f(x_1) \lambda(x_2) \rag \propto \theta(t_1-t_2) \hspace{0.3cm} \mbox{while} \hspace{0.3cm} \lag \lambda(x) \rag = 0 , \hspace{0.3cm} \lag \lambda(x_1) \lambda(x_2) \rag = 0 .
\label{resp2}
\eeq
The last two equalities in (\ref{resp2}) are obtained with $F=1$. The response function $R(x_1,x_2)$ is proportional to the Heaviside factor $\theta(t_1-t_2)$ because of causality: the field $f(x_1)$ at time $t_1$ only depends on the ``noise'' $\etai(x_2)$ at earlier times $t_2 \leq t_1$. Besides, the response $R$ obeys the constraints:
\beq
\mbox{for} \hspace{0.1cm} t_1 \rightarrow t_2^+ : \hspace{0.1cm} R(x_1,x_2) \rightarrow \delta_D(\ux_1-\ux_2) , \hspace{0.4cm} \mbox{and for all $t_1 \geq t_2$} : \hspace{0.2cm} \int \d\bx_1\d\bp_1 \; R(x_1,x_2) = 1 ,
\label{resp3}
\eeq
where the last property translates the fact that $\int \d\bx\d\bp \; f$ is a constant of motion of the Vlasov equation. Since our system is an out-of-equilibrium dynamics there is no fluctuation-dissipation theorem: there is no linear relationship between the response $R$ and the correlation $G(x_1,x_2)=\lag f(x_1)f(x_2) \rag_c$.

\subsection{Approximation schemes}
\label{Approximation schemes}

The path-integrals (\ref{Sf}) and (\ref{Sfl}) provide a convenient approach to investigate the statistical properties of the system. However, usually one does not know how to compute such non-Gaussian path-integrals so that one must resort to various approximation schemes. A first approach is to apply a perturbative expansion by expanding the exponential of the non-Gaussian part of the action (this would give a series over powers of $\Ks$). Such a perturbative scheme was described in \cite{Val1}, based on an integral form of the Vlasov equation which directly relates $\df$ to the linear growing mode $\eta_L$ (and not to the initial conditions at a finite time $\ti$), where $\df=f-\delta_D(\bp)$ is the deviation of the phase-space density $f$ from the homogeneous Hubble flow.  This yields another action $S[\df]$ whose non-Gaussian terms may be expanded. However, as shown in \cite{Val1}, this procedure actually recovers the standard perturbative results derived from the hydrodynamical approach. Therefore, it is not sufficient for our purpose which is to devise a theoretical tool to investigate the non-linear regime. On the other hand, a non-perturbative scheme is provided by the Feynman variational method (e.g., \cite{Fey1}) based on the inequality:
\beq
\int [\d f] \; e^{-S[f]} \geq e^{-\lag S-S_0 \rag_0} \int [\d f] \; e^{-S_0[f]} \hspace{0.3cm} \mbox{with} \hspace{0.3cm} \lag S-S_0 \rag_0 = \frac{\int [\d f] \; (S[f]-S_0[f]) \; e^{-S_0[f]}}{\int [\d f] \; e^{-S_0[f]}} .
\label{Fey}
\eeq
This applies to any real actions $S$ and $S_0$. Then, by maximizing the right-hand side of the inequality (\ref{Fey}) over a class of Gaussian actions $S_0$ where all integrals can be computed one may obtain a useful approximation. One may try to apply this method to the real action $S[f]$ derived in eq.(\ref{Sf}) (note that $S[f,\lambda]$ in eq.(\ref{Sfl}) is complex). Unfortunately, this method cannot be easily applied to our system because the matrix $\Dim$ is actually divergent since $\Di$ is not invertible, as can be seen from eqs.(\ref{etaL1})-(\ref{Di1}). Thus the matrix $\Dim$ is actually regularized in the action $S[f]$ through an infinitesimal perturbation and we must make sure that such divergences disappear in the final results. This behaviour merely translates the fact that the random field $\etai$ is constrained to be of the form (\ref{Di1}). This also means that the path-integral (\ref{Sf}) only runs over phase-space distributions $f(\bx,\bp,t)$ which are solutions of the Vlasov equation (\ref{Bol2}) with initial conditions of the form (\ref{Di1}). Thus, in order to keep the quantity $\lag S-S_0 \rag_0$ finite in (\ref{Fey}) we must only consider trial actions $S_0$ which also select solutions to the Vlasov equation with similar initial conditions. Therefore, we have not gained much by using this method: using such an action $S_0$ is usually just as difficult as the original problem with the action $S$. Nevertheless, one might think to use a trial action $S_0$ which only ``authorizes'' spherical linear density fields. Indeed, the gravitational dynamics of spherical fields is more easily solved (or at least easier to investigate) so that one could hope to derive in this way some useful information. However, although the average $\lag S-S_0 \rag_0$ is well-behaved (since we integrate over exact solutions of the Vlasov equation) the path-integral in the right-hand side of the inequality (\ref{Fey}) now involves a divergent Determinant. This comes from the fact that by integrating over spherical fields only we no longer have the same number of degrees of freedom in the systems defined by $S$ or $S_0$. Therefore, we cannot derive any rigorous information from the restriction of the original problem to spherical fields only (see also Sect.4 and App.A in \cite{Val4}).

A third approach to path-integrals is to apply a steepest-descent expansion. Therefore, we may add a multiplicative parameter $N$ to the action $S$ and define the generating functional $Z_N[j,h]$:
\beq
Z_N[j,h] = \int [\d f] [\d \lambda] \; e^{N[ j.f + h.\lambda - S(f,\lambda)]} \hspace{0.3cm} \mbox{whence} \hspace{0.3cm} Z_N[j] = \int [\d f] \; e^{N[ j.f - \frac{1}{2} (\cO f - \Ks f^2 -\etabi) . \Dim . (\cO f - \Ks f^2 -\etabi) ]} .
\label{ZNsc1}
\eeq
As is well-known, functional derivatives of $Z_N$ with respect to the source fields $j$ and $h$ yield all many-body correlations. The actual case corresponds to $N=1$ but we may compute $Z_N$ through an expansion over powers of $1/N$ and next put $N=1$ into the results. This procedure is similar to the ``semiclassical'' (or loopwise) expansion over powers of $\hbar$ encountered in quantum field theory. Note that this scheme is very different from the standard perturbative expansion described above since we make no difference between the quadratic and higher-order terms of the action, as the factor $N$ multiplies the whole action $S$.

Finally, a fourth method often used in physics to investigate such systems is to generalize our problem to an $N-$field system and look again for an expansion over powers of $1/N$. Of course, there are many ways to generalize the action $S[f]$ to $N$ fields. In order to obtain good scaling properties with $N$, one can check that it is convenient to first write the partition function $Z$ defined by the action $S[f]$ derived in (\ref{Sf}) as:
\beq
Z =  \int [\d f] [\d \lambda] [\d \sigma] \; e^{ - \frac{1}{2} (\cO f-\etabi).\Dim.(\cO f-\etabi) + \lambda . (\Ks f^2-\sigma) + (\cO f-\etabi).\Dim.\sigma - \frac{1}{2} \sigma.\Dim.\sigma } ,
\label{ZN1}
\eeq
where we introduced the auxiliary fields $\lambda$ and $\sigma$, with $\lambda$ imaginary. Next, we generalize our system from one field $f$ to $N$ similar fields $f_a$ with $1\leq a\leq N$. Thus, we write the partition function $Z_N'$ as:
\beqa
Z_N' & = & \int \prod_{a=1}^{N} [\d f_a] [\d \lambda] [\d \sigma] \; e^{ - \frac{1}{2} \sum_a (\cO f_a-\etabi).\Dim.(\cO f_a-\etabi) + \lambda . \left( \sum_a \Ks f_a^2-\sigma \right) + \sum_a (\cO f_a-\etabi).\Dim.\sigma - \frac{1}{2} \sigma.\Dim.\sigma } \\ & = & \int \prod_{a=1}^{N} [\d f_a] [\d \lambda] \; e^{ - \frac{1}{2} \sum_a (\cO f_a-\etabi).\Dim.(\cO f_a-\etabi) + \lambda . \sum_a \Ks f_a^2 + \frac{1}{2} \left[\Dim.\sum_a (\cO f_a-\etabi) - \lambda\right] .\Di. \left[\Dim.\sum_b (\cO f_b-\etabi) - \lambda\right] } ,
\label{ZN2}
\eeqa
where we performed the Gaussian integration over the field $\sigma$ in the last term. Then,  in order to ensure that all terms contribute to the same order over $N$ to the action, with the fields $f_a$ and $\lambda$ of order unity, we add some factors $N$ in the exponent and we define the action
$S_N$ by:
\beqa
S_N[f_a,\lambda] & = & \frac{1}{2} \sum_a (\cO f_a-\etabi).\Dim.(\cO f_a-\etabi) - \frac{1}{2N} \left[ \sum_a (\cO f_a-\etabi) \right].\Dim.\left[ \sum_b (\cO f_b-\etabi) \right] \nonumber \\ & &  + \lambda . \sum_a (\cO f_a - \Ks f_a^2 - \etabi) - \frac{N}{2} \lambda.\Di.\lambda
\label{SN1}
\eeqa
Of course, for $N=1$ we recover the action which corresponds to the case of only one field $f$, which is the case of practical interest to us. One can easily show that the real part of the action $S_N[f_a,\lambda]$ is again positive. Note that the first two terms of $S_N$ cancel out for $N=1$, which shows that there are many ways to generalize to $N$ fields, some of which being more appropriate. In fact, these two terms are necessary for the path-integral $Z_N$ to be well-defined. Otherwise, the action $S_N$ would only depend on the two fields $\lambda$ and $\sigma=\sum_a (\cO f_a - \Ks f_a^2 - \etabi)$ which would lead to divergences and inconsistencies (the system would not have truly been generalized to an $(N+1)-$field problem). Note that the action $S_N$ scales as $S_N \sim N$ for large $N$ (with all fields of order unity, compare also with eq.(\ref{ZNsc1})).

\section{Direct steepest-descent method}
\label{Direct steepest-descent method}

\subsection{General formalism}
\label{General formalism}

In the following, we shall describe how to apply to gravitational clustering three large-$N$ methods which have been introduced in field theory. We shall consider the ``semiclassical'' eq.(\ref{ZNsc1}) which provides simpler expressions than the $N$-field generalization (\ref{SN1}). The latter actually provides the same results (for $N=1$) if we pay attention to the long-wavelength divergences discussed in Sect.\ref{Long-wavelength divergences}, see Sect.\ref{Infrared divergences}. Although the action $S[f,\lambda]$ involves two fields we shall first recall the application of large-$N$ expansions to the case of a single scalar field $\phi(x)$ governed by a cubic action $S[\phi]$:
\beq
Z[j] = e^{N W[j]} = \int [\d\phi] \; e^{N [j.\phi-S(\phi)]} \hspace{0.3cm} \mbox{with} \hspace{0.3cm} S[\phi] = S_1(x) \phi(x) + \frac{1}{2} S_2(x,y) \phi(x) \phi(y) + \frac{1}{3!} S_3(x,y,z) \phi(x) \phi(y) \phi(z)
\label{Zphi1}
\eeq
This corresponds to our case since the action $S[f,\lambda]$ is indeed cubic. This will yield simpler expressions in the intermediate steps and we shall directly translate the final results to our case where $\phi=(f,\lambda)$. Without loss of generality, we take the kernels $S_2$ and $S_3$ to be fully symmetric. As is well-known, successive derivatives of the functional $W[j]$ with respect to $j$ yield the many-body connected correlations of the field $\phi$.

Obviously, a first approach to handle the large-$N$ limit of eq.(\ref{Zphi1}) is to use a steepest-descent method (also called a semiclassical or loopwise expansion in the case of usual Quantum field theory with $\hbar=1/N$). Thus, as described for instance in \cite{Zinn1} we may compute $W[j]$ by expanding the action $S[\phi]$ around the saddle-point $\phis$ which yields up to order $1/N^2$:
\beqa
W[j] & = & j.\phis - S(\phis) - \frac{1}{2N} \Tr \ln \Gom + \frac{1}{N^2} S_3(x_1,x_2,x_3) S_3(y_1,y_2,y_3) \biggl [ \frac{1}{8} \Go(x_1,x_2) \Go(y_1,y_2) \Go(x_3,y_3) \nonumber \\ & & + \frac{1}{12} \Go(x_1,y_1) \Go(x_2,y_2) \Go(x_3,y_3) \biggl ] + \cO(1/N^3) ,
\label{WN1}
\eeqa
where we introduced the saddle-point $\phis(j)$ and the $\phis-$dependent propagator $\Go$ defined by:
\beq
\left. \frac{\delta S}{\delta \phi(x)} \right|_{\phis} = j(x) \hspace{0.3cm} \mbox{and} \hspace{0.3cm} \Gom(x,y;\phis) = \left. \frac{\delta^2 S}{\delta\phi(x) \delta\phi(y)} \right|_{\phis} = S_2 + S_3.\phis , \hspace{0.3cm} \mbox{whence} \hspace{0.3cm} \frac{\delta \Gom}{\delta \phis} = S_3  , \hspace{0.3cm} \frac{\delta^2 \Gom}{\delta \phis\delta \phis} = 0 .
\label{Go1}
\eeq
Next, we define the Legendre transform $\Gamma[\varphi]$ of the functional $W[j]$ by:
\beq
\Gamma[\varphi] = - W[j] + j . \varphi \hspace{0.3cm} \mbox{with} \hspace{0.3cm} \varphi(x) = \frac{\delta W}{\delta j(x)} \hspace{0.3cm} \mbox{whence} \hspace{0.3cm} j(x) = \frac{\delta \Gamma}{\delta \varphi(x)} .
\label{Leg1}
\eeq
As is well-known, the mean $\phib$ and the two-point connected correlation $G$ of the field $\phi$ can be obtained from $\Gamma[\varphi]$ through the variational equations:
\beq
\phib = \varphi \hspace{0.2cm} \mbox{at the saddle-point of $\Gamma$:} \hspace{0.2cm} \left. \frac{\delta \Gamma}{\delta \varphi} \right|_{\phib} = 0 ,  \hspace{0.2cm} \mbox{and} \hspace{0.2cm} G = N \lag \phi(x_1) \phi(x_2) \rag_c = \frac{\delta^2 W}{\delta j \delta j} = \left( \frac{\delta^2 \Gamma}{\delta\varphi \delta\varphi} \right)^{-1} \hspace{0.2cm} \mbox{at} \hspace{0.2cm} \varphi=\phib .
\label{phib1}
\eeq
These relations are obtained from eq.(\ref{Leg1}), after we notice that the physical averages are obtained when the external source is zero: $j=0$. Note that for large $N$ the actual two-point connected correlation $\lag \phi \phi \rag_c = G/N$ vanishes. This is natural since from eq.(\ref{Zphi1}) one indeed expects the field $\phi$ to be ``frozen'' to the minimum $\phis$ of the action. Next, from the expansion (\ref{WN1}) and the definition (\ref{Leg1}) we obtain the $1/N$ expansion of $\Gamma[\varphi]$. This yields (e.g., \cite{Zinn1}):
\beq
\Gamma[\varphi] = S(\varphi) + \frac{1}{2N} \Tr \ln \Gom - \frac{1}{12 N^2} S_3(x_1,x_2,x_3) S_3(y_1,y_2,y_3) \Go(x_1,y_1) \Go(x_2,y_2) \Go(x_3,y_3) + \cO(1/N^3) ,
\label{GamN1}
\eeq
where $\Go(x,y;\varphi)$ is defined as in eq.(\ref{Go1}) but taken at the point $\varphi$. Of course, one can check in eq.(\ref{GamN1}) that only one line irreducible (1PI) diagrams\footnote{A graph is said to be ``one line irreducible'', or ``1-particle irreducible'' (1PI) in the language of Particle Physics, if it cannot be disconnected by cutting only one line. Otherwise, it is ``1-particle reducible''.} appear in the expansion of $\Gamma[\varphi]$. Besides, the expansion over powers of $1/N$ is actually a loopwise expansion and in  eq.(\ref{GamN1}) we have obtained the one-loop and two-loop diagrams. Next, from eq.(\ref{GamN1}) and eq.(\ref{phib1}) we obtain the first two correlations $\phib$ and $G$ at the corresponding order over $1/N$. Firstly, the saddle-point condition reads at order $1/N^2$:
\beq
\frac{\delta S}{\delta\varphi(x)} + \frac{1}{2N} \Tr \left[ \frac{\delta \Gom}{\delta \varphi(x)} . \Go \right] + \frac{1}{4 N^2} S_3 S_3 \left( \Go . \frac{\delta \Gom}{\delta \varphi(x)} . \Go \right) \Go \Go = 0 \hspace{0.4cm} \mbox{ at the point } \hspace{0.4cm} \varphi=\phib ,
\label{phib2}
\eeq
where we did not write the explicit arguments to shorten the expression. Secondly, eq.(\ref{phib1}) yields for the two-point correlation $G$:
\beq
G^{-1} = \Gom - \Sigma ,
\label{G1}
\eeq
where we introduced the self-energy $\Sigma$ given at order $1/N^2$ by:
\beqa
\Sigma & = & \frac{1}{2N} \Tr \left[ \frac{\delta \Gom}{\delta \varphi} . \Go . \frac{\delta \Gom}{\delta \varphi} . \Go \right] + \frac{1}{2 N^2} S_3 S_3 \left( \Go . \frac{\delta \Gom}{\delta \varphi} . \Go . \frac{\delta \Gom}{\delta \varphi} . \Go \right) \Go \Go \nonumber \\ & & + \frac{1}{2 N^2} S_3 S_3 \left( \Go . \frac{\delta \Gom}{\delta \varphi} . \Go \right) \left( \Go . \frac{\delta \Gom}{\delta \varphi} . \Go \right) \Go .
\label{Sigma1}
\eeqa
To derive these equations we used the fact that $\delta^2 \Gom/\delta\varphi\delta\varphi=0$ since the action $S[\phi]$ is cubic, see eq.(\ref{Go1}). Thus, once we have obtained the generating functional of the proper vertices $\Gamma[\varphi]$ up to the required order over $1/N$, as in eq.(\ref{GamN1}), we derive the mean $\phib$ and the two-point correlation $G$ from eq.(\ref{phib1}), as in eqs.(\ref{phib2})-(\ref{Sigma1}).

\subsection{Application to gravitational clustering}
\label{Application to gravitational clustering}

The usual application of this method to a system with $N$ fields, as in eq.(\ref{SN1}) (or in usual quantum field theories), is to first integrate over the $N$ fields $f_a$ since the action $S_N$ is indeed Gaussian over $f_a$. Then, the partition function $Z_N$ is given by a path-integral over the one field $\lambda$ with a new action which contains an explicit $N$-dependence. Therefore, we recover an expression which is similar to eq.(\ref{Zphi1}), to which we can apply the method described above. However, we cannot apply this procedure to our case since $\lambda.\Ks$ is not invertible (indeed $\lambda.\Ks f^2=0$ for all fields $f$ such that $f(\bk,\bw=0,t)=0$, see eq.(\ref{Bol2})). Therefore, integrating over the fields $f_a$ is not well suited to our problem since it would introduce singular terms for $N=1$. Nevertheless, we can still apply the method described above to the actions given in eqs.(\ref{ZNsc1})-(\ref{SN1}). Thus, in the case of the gravitational dynamics described by the action (\ref{ZNsc1}) we write:
\beq
\phi = \left( \bea{c} f \\ \lambda \ea \right) , \hspace{0.3cm} \phib = \left( \bea{c} \fb \\ \lambdab=0 \ea \right) , \hspace{0.3cm} \cGm_{0;ij}(x,y;\phib) = \frac{\delta^2 S[\phib]}{\delta\phib_i(x) \delta\phib_j(y)} \hspace{0.4cm} \mbox{and} \hspace{0.4cm} \cG_0 = (\cG_{0;ij}) = \left( \bea{lr} G_{0} & R_0 \\ \tR_0 & 0 \ea \right) ,
\label{phiphib1}
\eeq
where we note $x=(\bx,\bp,t)$, the transposed matrix $\tR_0(x,y)=R_0(y,x)$, and we used eq.(\ref{resp2}). Indeed, from eq.(\ref{ZNsc1}) we can see that $N$ can be absorbed in the normalization of the field $\lambda$ and the matrix $\Di$, hence the vanishing of the first two moments of $\lambda$ holds for any $N$ and is recovered at any order of the expansion over $1/N$. This can also be checked explicitly. Then, from eq.(\ref{ZNsc1}) we obtain for the matrix propagator $\cG_0(x,y;\phib)$:
\beq
\cGm_{0}(x,y;\phib) = \left( \bea{cc} - 2 \lambdab(u) \Ks(u;x,y) & \cO (y,x) - 2 \Ks(y;x,u)\fb(u) \\ \cO (x,y) - 2 \Ks(x;y,u)\fb(u) & - \Di(x,y) \ea \right) .
\label{G0m1}
\eeq
From the definition of the inverse we have: $\delta_{ij} \delta_D(x-y) = \cGm_{0;ik}(x,z).\cG_{0;kj}(z,y)$. Substituting the expressions (\ref{phiphib1}) and (\ref{G0m1}), with $\lambdab=0$, we obtain:
\beq
\cO(x,z) G_0(z,y) = 2 \Ks(x;z,u) \fb(u) G_0(z,y) ,
\label{G0N1}
\eeq
and:
\beq
R_0(x,z) \cO(z,y) = \delta_D(x-y) + 2 R_0(x,z) \Ks(z;y,u) \fb(u) , \hspace{0.3cm} \cO(x,z) R_0(z,y) = \delta_D(x-y) + 2 \Ks(x;z,u) \fb(u) R_0(z,y) .
\label{R0N1}
\eeq
Here we used $\tR_0(x,y)=R_0(y,x)$ and we took the limit $\ti \rightarrow 0$ so that $\Di \rightarrow 0$, see eq.(\ref{etaL1}). Indeed, the inverse $\Dim$ does not appear in these equations so that the limit $\ti \rightarrow 0$ is straightforward, which was not the case for the Feynman variational method (\ref{Fey}). The two eqs.(\ref{R0N1}) are redundant but for numerical purposes it can be useful to make sure they are both satisfied.  Thus, eqs.(\ref{G0N1})-(\ref{R0N1}) yield the auxiliary propagators $G_0$ and $R_0$. Next, we write the actual two-point correlation matrix $\cG$ and the self-energy $\self$ as (with $\Sigma(x_1,x_2) \propto \theta(t_1-t_2)$):
\beq
\cG = \left( \bea{lr} G & R \\ \tR & 0 \ea \right) \hspace{0.3cm} \mbox{with} \hspace{0.3cm} G(x,y) = N \lag f(x) f(y) \rag_c , \hspace{0.2cm} R(x,y) = N \lag f(x) \lambda(y) \rag_c , \hspace{0.3cm} \mbox{and} \hspace{0.3cm} \self = \left( \bea{lr} 0 & \tSigma \\ \Sigma & \Pi \ea \right) .
\label{Gij1}
\eeq
Then, convoluting from the right with $\cG$ the Schwinger-Dyson eq.(\ref{G1}), $\cGm=\cGm_0 - \self$, we obtain:
\beq
\cO(x,z) G(z,y) = 2 \Ks(x;z,u) \fb(u) G(z,y) + \Sigma(x,z) G(z,y) + \Pi(x,z) \tR(z,y)
\label{GN1}
\eeq
\beq
R(x,z) \cO(z,y) = \delta_D(x-y) + 2 R(x,z) \Ks(z;y,u) \fb(u) + R(x,z) \Sigma(z,y)
\label{tRN1}
\eeq
\beq
\cO(x,z) R(z,y) = \delta_D(x-y) + 2 \Ks(x;z,u) \fb(u) R(z,y) + \Sigma(x,z) R(z,y)
\label{RN1}
\eeq
Here we have again taken the limit $\ti \rightarrow 0$. Note that eqs.(\ref{GN1})-(\ref{RN1}) are exact. To complete our system we simply need to add the saddle-point condition (\ref{phib1}) and the expression of the self-energy $\self$. One can easily see that at tree-order for the generating functional $\Gamma[\varphi]$ we merely recover the linear regime (\ref{etaL1}). Therefore, we must go at least up to one-loop order to make progress. From eq.(\ref{GamN1}) this yields $\Gamma=S+\frac{1}{2N} \Tr\ln\cGm_0$. Then, the saddle-point condition (\ref{phib1}) now reads:
\beq
\frac{\delta \Gamma}{\delta \lambdab(x)}=0 : \hspace{0.3cm} \cO(x,y) \fb(y) = \frac{1}{N} \Ks(x;y,z) G_0(y,z) \hspace{0.5cm} \mbox{using} \hspace{0.3cm} \Ks \fb^2 = 0 \hspace{0.3cm} \mbox{with} \hspace{0.3cm} \fb(\bk,\bw=0,t) = \frac{\delta_D(\bk)}{(2\pi)^3} .
\label{fb1}
\eeq
Here we used eq.(\ref{kc1}). The second condition $\delta \Gamma/\delta \fb$ is trivially satisfied. Finally, the self-energy is given at one-loop order by the first term in the r.h.s. in eq.(\ref{Sigma1}) which yields:
\beq
\Sigma(x,y) = \frac{4}{N} \Ks(x;x_1,x_2) \Ks(z;y,z_2) R_0(x_1,z) G_0(x_2,z_2)
\label{Sigma1loop1}
\eeq
\beq
\Pi(x,y) = \frac{2}{N} \Ks(x;x_1,x_2) \Ks(y;y_1,y_2) G_0(x_1,y_1) G_0(x_2,y_2)
\label{Pi1loop1}
\eeq
We can check in eq.(\ref{Sigma1loop1}) that $\Sigma(x_1,x_2) \propto \theta(t_1-t_2)$. Thus, together with the boundary conditions set by the linear regime (in the limit $t \rightarrow 0$), eqs.(\ref{G0N1})-(\ref{Pi1loop1}), provide the mean $\fb$, the two-point correlation $G$ and the response $R$, up to one-loop order within the direct steepest-descent method. We simply need to solve these equations after setting $N=1$. One can also compute higher-order correlations through this method. Therefore, we have shown that despite the singular behaviour of the action $S[f]$ in eq.(\ref{Sf}), which prevented the use of the Feynman variational method (\ref{Fey}), we can apply the large$-N$ expansion given by the direct steepest-descent method. Moreover, it is clear from eq.(\ref{ZNsc1}) that all symmetries of the problem remain valid for any $N$ and hold at any order of the expansion over $1/N$. In particular, we recover the exact properties (\ref{resp2})-(\ref{resp3}).

\section{1PI effective action}
\label{1PI Effective action}

Although the direct steepest-descent method described in the previous section can provide some useful insight into the physics at work, it has been noticed in several cases that it could also fail completely. Thus, it may exhibit secular terms and high-order corrections may blow up with time (e.g., \cite{Mih1}). This has led to a shift of interest towards other approaches to large-$N$ expansions. We shall present in Sect.\ref{2PI effective action} such a scheme which has already been shown to provide a significant improvement over the direct steepest-descent method in several cases, where it was seen to be free of secular terms (\cite{Mih1}) or to exhibit the looked-for relaxation towards thermal equilibrium (\cite{Berg1}). However, we shall first present an alternative approach based on the ``1PI effective action'' $\Gamma[\varphi]$ defined in eq.(\ref{Leg1}). The method described in the previous section was to firstly derive a series expansion over $1/N$ for $W[j]$, secondly obtain through a Legendre transform an expansion for $\Gamma[\varphi]$ and thirdly compute the means $\phib$ and $G$ from this expression of the 1PI effective action. An alternative approach is to firstly write the exact equations of motion which determine $\Gamma[\varphi]$, secondly use a large-$N$ approximation for these equations and thirdly derive $\phib$ and $G$. That is, we directly apply the large-$N$ expansion to $\Gamma[\varphi]$ rather than $W[j]$. Since $W$ and $\Gamma$ are related through a non-linear Legendre transform, both procedures are not identical, although they must yield the same results up to the highest-order over $1/N$ of the required expansion (i.e. they only differ by higher-order terms). As described in standard textbooks (e.g., \cite{Itz1}) the Schwinger-Dyson equations can be obtained by noticing that the integral of a derivative vanishes:
\beq
\int [\d\phi] \; \frac{\delta}{\delta\phi(x)} \; e^{N[j.\phi-S(\phi)]} = 0 \hspace{0.4cm} \mbox{whence} \hspace{0.4cm} \left[ j(x) -  \frac{\delta S}{\delta\phi(x)} \left( \frac{1}{N} \frac{\delta}{\delta j} \right) \right] . Z[j] = 0 .
\label{Dys1}
\eeq
For the cubic action $S[\phi]$ defined in eq.(\ref{Zphi1}) this yields:
\beq
j(x) - S_1(x) - S_2(x,x_1).\frac{\delta W}{\delta j(x_1)} - \frac{1}{2} S_3(x,x_1,x_2) \left[ \frac{\delta W}{\delta j(x_1)} \frac{\delta W}{\delta j(x_2)} + \frac{1}{N} \frac{\delta^2 W}{\delta j(x_1)\delta j(x_2)} \right] = 0.
\label{Dys11}
\eeq
By taking successive derivatives over $j$ at the point $j=0$ of eq.(\ref{Dys11}) we would obtain a hierarchy of equations between the many-body correlations $G_q$. In this way we would recover the standard BBGKY hierarchy (\cite{Peeb1}). However, here we introduce the 1PI effective action defined in eq.(\ref{Leg1}) and we write eq.(\ref{Dys11}) in terms of $\Gamma[\varphi]$. This reads:
\beq
\frac{\delta\Gamma}{\delta\varphi} - \frac{\delta S}{\delta\varphi} - \frac{1}{2N} S_3.G = 0 \hspace{0.4cm} \mbox{with} \hspace{0.4cm} G= \frac{\delta^2 W}{\delta j \delta j} = \left( \frac{\delta^2 \Gamma}{\delta\varphi \delta\varphi} \right)^{-1} .
\label{Dys2}
\eeq
As seen from eq.(\ref{phib1}) $G/N$ is actually the two-point correlation of the field $\phi$ at the point $\varphi=\phib$. The equation (\ref{Dys2}) is exact. We only used the fact that the action $S[\phi]$ is cubic as given in eq.(\ref{Zphi1}). From eq.(\ref{phib1}) we note that at the saddle-point $\phib$ of $\Gamma$ we have the relation:
\beq
\frac{\delta S}{\delta\varphi} + \frac{1}{2N} S_3.G = 0  \hspace{0.4cm} \mbox{at the saddle-point} \hspace{0.4cm} \varphi=\phib \hspace{0.4cm} \mbox{and with} \hspace{0.4cm} G=N\lag\phi\phi\rag_c.
\label{phib3}
\eeq
The eq.(\ref{Dys2}) is an intricate non-linear equation for the functional $\Gamma[\varphi]$ which we do not know how to solve. However, by taking successive derivatives of eq.(\ref{Dys2}) with respect to $\varphi$ we can obtain a useful hierarchy of equations. A first derivative of eq.(\ref{Dys2}) yields:
\beq
\Gm = \Gom - \Sigma  \hspace{0.4cm} \mbox{with} \hspace{0.4cm} \Gom(x,y;\varphi) = \frac{\delta^2 S(\varphi)}{\delta\varphi(x) \delta\varphi(y)} \hspace{0.4cm} \mbox{and} \hspace{0.4cm} \Sigma(x,y) = - \frac{1}{2N} S_3(x,.,.) \frac{\delta G(.,.)}{\delta\varphi(y)} .
\label{Dys3}
\eeq
The propagator $\Go$ and the self-energy $\Sigma$ are defined as in eq.(\ref{Go1}) and eq.(\ref{G1}). However, the last expression for $\Sigma$ in eq.(\ref{Dys3}) is exact while eq.(\ref{Sigma1}) is only its expansion up to order $1/N^2$. Introducing the three-point vertex $\Gamma_3$ we can also write:
\beq
\Sigma(x,y) = \frac{1}{2N} S_3(x,u_1,u_2) G(u_1,u_3) \Gamma_3(y,u_3,u_4) G(u_4,u_2) \hspace{0.4cm} \mbox{with} \hspace{0.4cm} \Gamma_3 = \frac{\delta^3\Gamma}{\delta\varphi\delta\varphi\delta\varphi} = \frac{\delta \Gm}{\delta\varphi} .
\label{Sigma2}
\eeq
Next, taking a second derivative of eq.(\ref{Dys2}) we get:
\beq
\Gamma_3 - S_3 - \frac{1}{N} S_3 . G . \Gamma_3 . G . \Gamma_3 . G + \frac{1}{2N} S_3 . G . \Gamma_4 . G = 0 \hspace{0.4cm} \mbox{with} \hspace{0.4cm} \Gamma_4 = \frac{\delta^4\Gamma}{\delta\varphi\delta\varphi\delta\varphi\delta\varphi} .
\label{Gamma31}
\eeq
Obviously, by taking successive derivatives of the equation of motion (\ref{Dys2}) we obtain an infinite hierarchy of equations which link the successive proper correlations $\Gamma_q$. However, because of the factor $1/N$ in eq.(\ref{Dys2}) higher-order vertices $\Gamma_q$ contribute to the self-energy $\Sigma$ at a higher order over $1/N$. Therefore, we can close this hierarchy by looking for an expression of $\Sigma$ up to a given order over $1/N$. Moreover, we note that since the action $S[\phi]$ is cubic there is no ``bare contribution'' $S_q$ to the $q-$point vertex $\Gamma_q$ for $q\geq 4$ so that $\Gamma_q \sim 1/N$ for $q\geq 4$. Note that this is consistent with eq.(\ref{GamN1}). Hence we obtain at order $1/N$ for $\Gamma_3$:
\beq
\Gamma_3(x,y,z) = S_3(x,y,z) + \frac{1}{N} S_3(x,u_1,u_2) G(u_1,u_3) \Gamma_3(y,u_3,u_4) G(u_4,u_5) \Gamma_3(z,u_5,u_6) G(u_6,u_2) + \cO(1/N^2) .
\label{Gamma32}
\eeq
Thus, we obtain a closed system and we can solve for $\phib$, $G$ and $\Gamma_3$ through eqs.(\ref{phib3})-(\ref{Gamma32}). Of course, the equations derived in this section agree with those obtained in Sect.~\ref{Direct steepest-descent method} up to order $1/N^2$ since they were both derived through $1/N$ expansions up to this order. However, they differ by higher-order terms. This is obvious from the fact that $\Sigma$ now depends on the exact propagator $G$ so that solving exactly the equation (\ref{Dys3}), with $\Sigma$ obtained from eq.(\ref{Gamma32}), will generate contributions to $G$ at all orders over $1/N$. In other words, by going from the method presented in Sect.\ref{Direct steepest-descent method} to the present method we have made some infinite partial resummations. This also means that we must solve these equations exactly. Indeed, there is no point to solve them as a perturbative series over $1/N$ since this would give back the results obtained from the direct steepest-descent method described in Sect.\ref{Direct steepest-descent method}.

Since both this 1PI effective action method and the former steepest-descent approach are derived up to the same order over $1/N$ one might naively expect their accuracy to be of similar order. Moreover, although the 1PI effective action scheme includes partial resummations with respect to the other method, it is not obvious a priori that this should improve the final results. Indeed, we still miss some of the higher-order terms and we can easily imagine peculiar cases where a resummation of some badly chosen diagrams would actually dramatically worsen the results (see also Sect.\ref{Infrared divergences} below). However, these new equations of motion for $\fb$ and $G$ were seen in several instances to yield a significant improvement over the steepest-descent scheme (e.g., \cite{Mih1}). This can be understood in a more generic context from the fact that the self-energy $\Sigma$ now depends on the actual correlation $G$ so that the Schwinger-Dyson equation $\Gm=\Gom-\Sigma$ is now an implicit equation for $G$. Since implicit schemes are usually more stable than explicit ones and satisfy physical constraints it is not surprising that the 1PI effective action approach should perform better (a familiar example is provided by the numerical integration of ordinary differential equations). This can also be analyzed in the light of virial scalings for the case of the gravitational dynamics, see the discussion in Sect.\ref{Conclusion}.

Finally, we note that these equations were also used in \cite{Mih1} to study the quantum roll. Going to order $1/N$ for $\Sigma$, they called this approximation the ``bare vertex approximation'' since at this order we simply have $\Gamma_3=S_3$. Here we have seen that one can actually go to any order over $1/N$ by including all vertices $\Gamma_q$ up to the required order. The last vertex is indeed given by its ``bare contribution'' $S_q$, which in our case yields $\Gamma_q=0$ if we close at a level $q \geq 4$. Thus, we see that we have actually closed the infinite hierarchy of equations for the correlations $\Gamma_q$ by neglecting the high-order proper correlations $\Gamma_q$ beyond some finite order. This is somewhat similar to the standard closure of the BBGKY hierarchy. However, the truncation of the proper correlations $\Gamma_q$ is much more powerful since it should still be able to handle the non-linear regime. This is not surprising since, as is well-known, as soon as $\Gamma_3 \neq 0$ one obtains a non-zero estimate for all many-body correlations $G_q$. Therefore, working with $\Gamma[\varphi]$ allows us to investigate powerful closure schemes which may still handle the strongly non-Gaussian regime. Note also that this 1PI effective action method provides an estimate for all many-body correlations $G_q$ (since from $\Gamma$ we can derive $W$) although in this paper we restrict ourselves to $\fb$ and $G$.

\section{2PI effective action}
\label{2PI effective action}

\subsection{General formalism}
\label{General formalism2PI
}

In the 1PI effective action approach developed in Sect.\ref{1PI Effective action} one needs to study higher-order proper correlations $\Gamma_q$ as one pushes the method to higher orders over $1/N$. This is not very convenient for numerical purposes since it implies that one needs to solve implicit non-linear equations for quantities which depend on an increasingly large number of variables. This is especially critical in our case where the coordinate $x=(\bx,\bp,t)$ is actually seven-dimensional so that one very quickly goes beyond computer power. Therefore, we introduce in this section another large$-N$ method which is likely to be more powerful to study large-scale structure formation. It only involves the one and two-point correlations $\phib$ and $G$, whatever the order over $1/N$, but it again yields an implicit equation for $G$ (which is a key feature to handle the non-linear regime). This can be done by introducing higher-order Legendre transforms (e.g., \cite{Cyr1}), which are a direct generalization of the 1PI effective action $\Gamma$ used in Sect.\ref{1PI Effective action}. Thus, following \cite{CJT} we define the double generating functionals $Z[j,M]$ and $W[j,M]$ by:
\beq
Z[j,M] = e^{N W[j,M]} = \int [\d\phi] \; e^{N [j.\phi + \frac{1}{2} \phi.M.\phi - S(\phi)]} ,
\label{Z2phi1}
\eeq
for any test-field $j(x)$ and symmetric matrix $M(x_1,x_2)$. Next, we define the double Legendre transform $\Gamma[\varphi,G]$ as:
\beq
\Gamma[\varphi,G] = - W[j,M] + j.\varphi + \frac{1}{2} \varphi.M.\varphi + \frac{1}{2N} \Tr [G.M] , \hspace{0.4cm} \mbox{with} \hspace{0.3cm} \frac{\delta W}{\delta j}= \varphi \hspace{0.3cm} \mbox{and} \hspace{0.3cm} \frac{\delta W}{\delta M}= \frac{1}{2} \left[ \varphi \varphi + \frac{1}{N} G \right] .
\label{Gam2def}
\eeq
This is the generalization of eqs.(\ref{Zphi1}), (\ref{Leg1}). In particular, as in eq.(\ref{Leg1}) we also have:
\beq
\frac{\delta \Gamma}{\delta \varphi}= j+M.\varphi \hspace{0.3cm} \mbox{and} \hspace{0.3cm} \frac{\delta \Gamma}{\delta G}= \frac{1}{2N} M , \hspace{0.4cm} \mbox{while} \hspace{0.3cm} \Gamma[\varphi] = \Gamma[\varphi,G] \hspace{0.3cm} \mbox{at the saddle-point G such that } \hspace{0.3cm} \frac{\delta \Gamma}{\delta G}= 0.
\label{invers1}
\eeq
The last relation in eq.(\ref{invers1}) gives the 1PI effective action $\Gamma[\varphi]$ used in Sect.\ref{1PI Effective action} in terms of the new double Legendre transform $\Gamma[\varphi,G]$. Besides, as in eq.(\ref{phib1}) the mean $\phib$ and the two-point correlation $\lag \phi \phi \rag_c$ are obtained from the variational principle:
\beq
\phib=\varphi \hspace{0.3cm} \mbox{and} \hspace{0.3cm} \lag \phi \phi \rag_c = \frac{1}{N} G \hspace{0.3cm} \mbox{at the saddle-point } (\varphi,G) \mbox{ such that } \hspace{0.3cm} \frac{\delta \Gamma}{\delta \varphi}=0 , \hspace{0.3cm} \frac{\delta \Gamma}{\delta G}= 0.
\label{corr2}
\eeq
Moreover, while $\Gamma[\varphi]$ defined in Sect.\ref{Direct steepest-descent method} was the generating functional of 1PI diagrams with ``bare propagator'' $S_2^{-1}$ the double Legendre transform $\Gamma[\varphi,G]$ is the generating functional in $\varphi$ of two-line irreducible (2PI) diagrams\footnote{A graph is said to be ``two line irreducible'', or ``2-particle irreducible'' (2PI), if it cannot be disconnected by cutting only two lines. Otherwise, it is ``2-particle reducible''.} with propagator $G$ (e.g., \cite{Cyr2}, \cite{Vas1}, see also App.\ref{Reduced 2PI effective action}). Then, following \cite{CJT} it is convenient to write $\Gamma[\varphi,G]$ in the form:
\beq
\Gamma[\varphi,G] = S[\varphi] + \frac{1}{2N} \Tr \left[ \Gom.G -1 \right] - \frac{1}{2N} \Tr \ln \left( \Dm . G \right) - \frac{1}{N} \tGam[\varphi,G] \hspace{0.3cm} \mbox{with} \hspace{0.3cm} \Dm= S_2 , \hspace{0.3cm} \Gom= \frac{\delta^2 S}{\delta\varphi \delta\varphi} .
\label{tGam1}
\eeq 
Note that the propagator $\Go$ is defined as in eqs.(\ref{Go1}) and (\ref{Dys3}) while $\Delta$ is the ``bare propagator'' of the action $S[\phi]$. Thus, eq.(\ref{tGam1}) defines the new generating functional $\tGam$. Then, as shown in \cite{CJT} the reduced 2PI effective action $\tGam[\varphi,G]$ is actually the sum of 2PI vacuum graphs of a theory defined by the action $\tS[\phi]$ given by:
\beq
\tS[\phi] = \frac{N}{2} \phi.\Gm.\phi + N \Sint[\phi;\varphi] \hspace{0.3cm} \mbox{where} \hspace{0.3cm} \Sint[\phi;\varphi] = \mbox{ cubic and higher-order terms over } \phi \mbox{ of } S[\varphi+\phi] .
\label{tS2}
\eeq
As seen in eq.(\ref{tS2}), in general the action $\tS[\phi]$ depends parametrically on the field $\varphi$, whence the dependence on $\varphi$ of the functional $\tGam$. However, in the case of a cubic action as (\ref{Zphi1}), we see that the interaction part $\Sint[\phi;\varphi]$ introduced in eq.(\ref{tS2}) does not depend on $\varphi$. Therefore, the functional $\tGam$ does not depend on $\varphi$ either and we have:
\beq
\mbox{for a cubic action $S$:} \hspace{0.4cm} \tGam[G] = \mbox{ 2PI vacuum diagrams of the action } \tS[\phi] =  \frac{N}{2} \phi.\Gm.\phi + \frac{N}{3!} S_3 \phi^3 .
\label{tGam2}
\eeq
We give a simple proof of the result (\ref{tGam2}) in App.\ref{Reduced 2PI effective action}, following the analysis used in \cite{Vas1} to show that the full effective action $\Gamma[\varphi,G]$ only contains 2PI diagrams. Indeed, the case of a cubic action $S[\phi]$ where the reduced 2PI effective action $\tGam$ only depends on $G$ is much simpler than the general case. Since this is actually the relevant case for gravitational dynamics and the formalism associated with the double Legendre transform $\Gamma[\varphi,G]$ is not so widely used, App.\ref{Reduced 2PI effective action} makes this article self-contained. In particular, the analysis presented in App.\ref{Reduced 2PI effective action} should be sufficient for readers who are only interested in the application of this method to the problem we investigate here: the formation of large-scale structures through gravitational instability. The result (\ref{tGam2}) provides a loopwise expansion of $\tGam[G]$ over powers of $1/N$. From this expression for $\Gamma$, we can derive the mean $\phib$ and the correlation $G$ through the variational equations (\ref{corr2}). Besides, from $\Gamma$ we can also obtain $W$ and all correlations at any order. Of course, we can see that the saddle-point condition (\ref{corr2}) for $G$ yields a non-linear implicit equation for $G$ (since $\tGam$ is a functional of $G$) as we wished. More precisely, let us write eq.(\ref{corr2}) using eq.(\ref{tGam1}). This reads:
\beq
\frac{\delta \Gamma}{\delta \varphi}=0 : \hspace{0.3cm} \frac{\delta S}{\delta\varphi} + \frac{1}{2N} \Tr \left[ \frac{\delta \Gom}{\delta\varphi}.G \right] = 0 \hspace{0.3cm} \mbox{whence} \hspace{0.3cm} \frac{\delta S}{\delta\varphi} + \frac{1}{2N} S_3.G = 0 .
\label{varphi1}
\eeq
Thus, we recover the exact equation (\ref{phib3}). Indeed, so far we have made no approximation. We only used the fact that the action $S[\phi]$ is cubic. Next, the saddle-point condition for $G$ writes:
\beq
\frac{\delta \Gamma}{\delta G}=0 : \hspace{0.3cm} \Gm = \Gom - \Sigma \hspace{0.3cm} \mbox{with} \hspace{0.3cm} \Sigma = 2 \frac{\delta \tGam}{\delta G} .
\label{varG1}
\eeq
This equation gives the exact expression of the self-energy $\Sigma$ in terms of $\tGam$. Note that eq.(\ref{varG1}) clearly implies that $\tGam$ must be two-line irreducible (2PI) since, as is well known, only 1PI graphs contribute to $\Sigma$, see also \cite{Berg1}.

\subsection{Comparison with the 1PI effective action approach and the BBGKY hierarchy}
\label{Comparison with the 1PI effective action approach}

In order to compare this method to the 1PI effective action scheme described in Sect.\ref{1PI Effective action}, let us compute the two-loop and three-loop contributions to $\tGam$. This yields:
\beq
\tGam = \frac{1}{12N} S_3 S_3 G^3 + \frac{1}{24N^2} S_3 S_3 S_3 S_3 G^6 ,
\label{tGam3}
\eeq
which gives for the self-energy $\Sigma$:
\beqa
\lefteqn{ \Sigma(x,y) = \frac{1}{2N} S_3(x,x_2,x_3) S_3(y,y_2,y_3) G(x_2,y_2) G(x_3,y_3) } \nonumber \\ & & + \frac{1}{2N^2} S_3(x,x_2,x_3) S_3(y,y_2,y_3) S_3(z_1,z_2,z_3) S_3(u_1,u_2,u_3) G(x_2,z_1) G(x_3,u_1) G(y_2,z_2) G(y_3,u_2) G(z_3,u_3) .
\label{Sigma3}
\eeqa
Then, we see that both methods are actually identical when we close the hierarchy in Sect.\ref{1PI Effective action} at order $1/N$ and we only keep the two-loop term in eq.(\ref{tGam3}). However, when we go to higher orders (here $1/N^2$ or three-loop) these two methods are no longer identical, but of course they still agree up to the order over $1/N$ up to which they were derived (i.e. they differ by higher-order terms). This is not surprising. Indeed, as described in Sect.\ref{1PI Effective action}, in the previous 1PI effective action method as we go to higher orders over $1/N$ we close the hierarchy of equations for the vertices $\Gamma_q$ at higher levels. Hence we include an increasingly large number of exact equations of motion for these correlations $\Gamma_q$ (hence for the connected correlations $G_q$). This can be a nice feature, but as we noticed above it is not very convenient for numerical purposes since it leads to non-linear implicit equations which involve higher-order correlations. By contrast, in the approach discussed here the basic unknowns are the field $\varphi$ and the correlation $G$, whatever the order over $1/N$. Indeed, although high-order diagrams for $\tGam$ can lead to lengthy expressions for the self-energy $\Sigma$ they only involve the explicit vertex $S_3$ and the two-point correlation $G$. Of course, the drawback is that the only exact equation of motion of the previous hierarchy which remains valid within this approach is the first relation (\ref{varphi1}). 

This also clearly shows how we could extend this method to include exactly some higher-order equations of the hierarchy. We simply need to work with higher-order Legendre transforms $\Gamma[\varphi,G,G_3,..,G_q]$ (see also \cite{Cal1}). Then, we would recover all exact BBGKY equations between the first few correlations $\phib,G,G_3,..,G_q$, up to order $q$. Of course, as in Sect.\ref{1PI Effective action}, the advantage of the method described here with respect to a simple truncation of the BBGKY hierarchy itself, is that higher-order correlations $G_{q+1},..,$ are not taken to be zero. Rather, any expression for the functional $\tGam$ (or $\Gamma$) defines a precise closure where higher-order correlations ($G_{q+1},..$) are expressed in terms of the lower-order correlations ($\phib,G,..,G_q$) which are explicitly taken into account in the Legendre transform $\Gamma$. Thus, by working with the double Legendre transform $\Gamma[\varphi,G]$ introduced in eq.(\ref{Gam2def}), we keep the exact BBGKY relation between $\phib$ and $G$ and we approximate higher-order correlations $G_3,..,$ by some functionals of $\phib$ and $G$. Of course, this also means that these higher-order correlations $G_q$ will no longer satisfy the exact BBGKY hierarchy. This is why the 2PI effective action method described here is not identical to the scheme presented in Sect.\ref{1PI Effective action}.

However, once we have an approximate 2PI effective action $\Gamma[\varphi,G]$, which also defines the usual 1PI effective action $\Gamma[\varphi]$ through eq.(\ref{invers1}), then all many-body correlations $G_q$ will satisfy the new BBGKY hierarchy associated with this new physical system (which we define by its 2PI effective action). Thus, the method described in this section is rather elegant as it keeps the variational structure (\ref{corr2}) and the Legendre relation (\ref{Gam2def}). Besides, in the context of Quantum field theories it can be shown that in some specific cases (\cite{Baym1}) some very important properties can be proved from the mere fact that the equations of motion can be derived through a variational principle as in eq.(\ref{varG1}). In our case, these consequences have no practical interest but it might happen that another similar effect comes into play (note indeed that the fact that the self-energy is the derivative of some functional $2\tGam[G]$ is a non-trivial constraint, especially when the field $\phi$ is actually a composite field as in our case).

\subsection{Equations of motion for gravitational clustering}
\label{Equations of motion for gravitational clustering}

\begin{figure}[htb]

\centerline{\epsfxsize=10 cm \epsfysize=3 cm \epsfbox{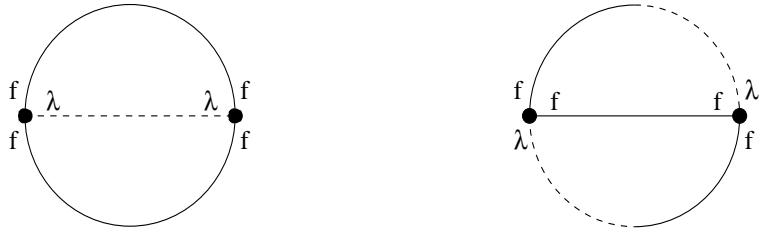}}

\caption{The two-loop 2PI vacuum diagrams of the reduced effective action $\tGam$. Each solid line stands for a propagator $G$, each dashed line for the propagator $D$ and each full-dashed line for the mixed propagator $R$ (cross-correlation). The indices $f,\lambda$ denote the $f-$field and $\lambda-$field sides of the propagators $G$, $R$ or $D$. These indices $f,\lambda$ also correspond to the legs of the three-leg vertex $\Ks$ (big dots).}
\label{fig2loopN}

\end{figure}

We now apply to gravitational clustering, described by eq.(\ref{ZNsc1}), this 2PI effective action method. Thus, eq.(\ref{varphi1}) reads:
\beq
\frac{\delta \Gamma}{\delta \lambdab(x)}=0 : \hspace{0.3cm} \cO(x,y) \fb(y) = \frac{1}{N} \Ks(x;y,z) G(y,z) ,
\label{Gamfb}
\eeq
while the second condition $\delta \Gamma/\delta \fb$ is again trivially satisfied. Note the difference between eq.(\ref{Gamfb}) and eq.(\ref{fb1}) obtained in the direct steepest-descent method. As explained above, eq.(\ref{Gamfb}) is actually exact. In fact, for $N=1$ it can be directly obtained by taking the average of the Vlasov equation (\ref{Bol3}) (with $\ti \rightarrow 0$). Next, the Schwinger-Dyson equation (\ref{varG1}) yields again eqs.(\ref{GN1})-(\ref{RN1}) as in Sect.\ref{Direct steepest-descent method}. However, the self-energy $\self$ is different and it is now derived from the functional $\tGam$, which is obtained through a loopwise expansion of the 2PI vacuum diagrams defined by the action $\tS$. The latter is given by: $\Sint[\phi] = - \lambda . \Ks f^2$ and we now have three propagators: $G, R$ and $D=N\lag \lambda \lambda \rag_c$. Since $D$ eventually vanishes it was not needed within the direct steepest-descent scheme presented in Sect.\ref{Direct steepest-descent method}. However, within the 2PI effective action method we must keep all terms in the intermediate steps. Then, the two-loop contribution to $\tGam$ is given by the two graphs displayed in Fig.\ref{fig2loopN}, which yields:
\beqa
\tGamtwo & = & \frac{1}{2N} \lag (-\lambda.\Ks f f) (-\lambda.\Ks f f) \rag_{\rm 2PI} \nonumber \\ & = & \frac{1}{N} \Ks(x;x_1,x_2) \Ks(y;y_1,y_2) \left[ D(x,y) G(x_1,y_1) G(x_2,y_2) + 2 \tR(x,y_1) G(x_2,y_2) R(x_1,y) \right] .
\label{Gam2loop1}
\eeqa
Next, from eq.(\ref{Gam2loop1}) and eq.(\ref{varG1}), which now reads $\self = (2 \delta \tGam/\delta \cG_{ij})$, we obtain the two-loop contribution to the self-energy. This yields:
\beq
\Sigmatwo(x,y) = \frac{4}{N} \Ks(x;x_1,x_2) \Ks(z;y,z_2) R(x_1,z) G(x_2,y_2)
\label{Sigma2loop1}
\eeq
\beq
\Pitwo(x,y) = \frac{2}{N} \Ks(x;x_1,x_2) \Ks(y;y_1,y_2) G(x_1,y_1) G(x_2,y_2)
\label{Pi2loop1}
\eeq
Moreover, we can check that $D$ and the (1,1) component of $\self$ vanish. Note again the difference between eqs.(\ref{Sigma2loop1})-(\ref{Pi2loop1}) and eqs.(\ref{Sigma1loop1})-(\ref{Pi1loop1}). By going to higher orders over $1/N$ (higher-loop diagrams) we would add higher powers over $\fb,G$ and $R$ to the self-energy. We obtain a finite number of terms because to each order over $1/N$ only corresponds a finite number of diagrams in the expansion of the reduced 2PI effective action $\tGam$. This nice feature arises because we kept the auxiliary field $\lambda$ throughout the calculation. Since the actions in eqs.(\ref{ZNsc1})-(\ref{SN1}) are quadratic over $\lambda$ we could have performed the Gaussian integration over $\lambda$. Then, we could apply the same formalism. However, the drawback of this approach is that to each order over $1/N$ would now correspond an infinite number of diagrams which we would need to resum (see \cite{Aarts1}). Thus, keeping the auxiliary field $\lambda$ allows us to perform these resummations in an automatic manner through the matrix $R$. Moreover, as shown in Sect.\ref{Derivation of the actions} the field $\lambda$ actually generates the response-functions of the system so that $R$ also has a physical interpretation as the mean first-order response function.

\subsection{Infrared divergences}
\label{Infrared divergences}

Although we applied the 2PI effective action method to the action $S[f,\lambda]$ given in eq.(\ref{ZNsc1}), we could also apply it to the $(N+1)-$field action (\ref{SN1}), looking for a symmetric solution. The analysis is somewhat more intricate because we now have four different propagators $G,\tG,R$ and $D$ ($\tG$ corresponds to $\lag f_a f_b \rag_c$ with $a \neq b$ while $G$ corresponds to $\lag f_a f_a \rag_c$). Moreover, these propagators now exhibit different scalings over $N$. Thus, there is no longer a one-to-one correspondence between the loop-order of a contribution to the reduced effective action $\tGam$ and the order over $1/N$ of the required correlation $G$. One needs to check explicitly by inspection of the equations of motion the order over $1/N$ required for each self-energy term to obtain the correlations $G,\tG,R$ and $D$ up to the looked-for order over $1/N$. In particular, within this framework the diagram to the right of Fig.\ref{fig2loopN} yields a subleading contribution as compared to the left diagram so that it would be disregarded. This implies that the frameworks (\ref{ZNsc1}) and (\ref{SN1}) are not equivalent for $N \neq 1$.

However, before we can apply any of these procedures (i.e. working with (\ref{ZNsc1}) or (\ref{SN1})) we must pay attention to the long-wavelength divergences which we discussed in Sect.\ref{Long-wavelength divergences}. We recalled that within the framework of standard perturbation theory these infrared divergences cancel out, at least at leading order (e.g., \cite{Jain1}). Obviously, we need to preserve this property in our approximation scheme. Otherwise, our results would be grossly inaccurate and useless (even though realistic power-spectra have $n \geq -1$ on scales $x \ga 10$ Mpc, this effect would prevent accurate results at smaller scales which are precisely those of interest). Therefore, we must make sure that we preserve these delicate cancellations between different diagrams. Thus, once we have selected all diagrams which are needed to reach the order over $1/N$ which we wish to obtain, we must add the sub-leading diagrams (if any) which contribute to higher orders over $1/N$ but provide the right ``counter-terms'' (at $N=1$) to the infrared divergences of the graphs we have already included.

The identification of the right diagrams to group together in order to cure these infrared problems is not obvious in the expansion over $1/N$ if we work with the $(N+1)-$field action (\ref{SN1}). Indeed, it is a completely unrelated point. Fortunately, this question is easily solved by working with the ``semiclassical expansion'' (\ref{ZNsc1}). Indeed, as seen for instance from the second expression in eq.(\ref{ZNsc1}) we see that the factor $N$ can be absorbed within the normalization of the correlation $\Di$. Therefore, the expansion over $1/N$ is actually an expansion over the amplitude of $\Di$, hence over powers of the linear power-spectrum $\PLo(k)$, see Sect.\ref{Initial conditions}. As a consequence, it shows the same cancellations of the infrared divergences as the usual perturbative approach for any $N$ and at any order over $1/N$. On the other hand, since the $(N+1)-$field approach does not affect the same powers of $1/N$ to the various diagrams it does not exhibit this infrared cancellation (except for $N=1$ if we sum all diagrams). Therefore, it is not well-suited to gravitational clustering studies\footnote{It is interesting to note that in a recent paper, studying for a $1+1$ dimensional $\phi^4$ field theory several large-$N$ schemes like those described in this article, \cite{Coop1} found that the accuracy of the numerical calculations as compared with the exact result at $N=1$ was better when one keeps both diagrams in Fig.\ref{fig2loopN}. We have seen that this is also the case here because of the infrared problems. One may wonder whether this is a general feature.}, unless we use the vertex given by the modified Vlasov equation (\ref{Bol6}) which automatically damps long-wavelength contributions. However, as noticed in Sect.\ref{Long-wavelength divergences}, this implies the loss of homogeneity. Hence, we can conclude that to investigate gravitational clustering we had better use the ``semiclassical'' expression (\ref{ZNsc1}) rather than the $(N+1)-$field generalization (\ref{SN1}).

\section{Perturbative expansion}
\label{Perturbative expansion}

Finally, we now solve eqs.(\ref{G0N1})-(\ref{Pi1loop1}) as a perturbative series over the linear power-spectrum $\PLo$. This actually boils down to the standard perturbative approach and exhibits the same ultraviolet divergences. Hence it cannot handle the non-linear regime but our purpose here is to check that the infrared divergences discussed in Sect.\ref{Infrared divergences} indeed disappear when we compute the equal-time correlation of the density field. In this section, we consider a critical-density universe and we normalize the time $t$ and the scale factor $a$ to unity at $z=0$. We also define a new time coordinate $\tau$ by:
\beq
\tau = \frac{1}{3}\ln (t/t_0) = \frac{1}{3}\ln t , \hspace{0.2cm}
\mbox{hence} \hspace{0.2cm} t=e^{3\tau} , \hspace{0.2cm} a = e^{2\tau}
\hspace{0.2cm} \mbox{and} \hspace{0.2cm} \rhob = \frac{1}{6 \pi \cG} .
\label{tau}
\eeq
Then, the linear operator $\cO$ and the vertex $K$ write in Fourier space:
\beq
\cO = \frac{\pl}{\pl \tau} - 3 e^{-\tau} \bk.\frac{\pl}{\pl \bw} , \hspace{0.3cm} K(x;x_1,x_2) = (2\pi)^3 2 e^{\tau} \delta_D(\tau_1-\tau) \delta_D(\tau_2-\tau) \delta_D(\bk_1+\bk_2-\bk) \delta_D(\bw_1) \delta_D(\bw_2-\bw) \frac{\bk_1.\bw_2}{k_1^2} ,
\label{O1}
\eeq
while from Sect.\ref{Initial conditions} the two-point correlation $\DL=\lag\eta_L(\bk_1,\bw_1,\tau_1)\eta_L(\bk_2,\bw_2,\tau_2)\rag_c$ of the linear growing mode reads:
 \beq
\DL(\bk_1,\bw_1,\tau_1;\bk_2,\bw_2,\tau_2) = \frac{1}{(2\pi)^6} \; \delta_D(\bk_1+\bk_2) \; \PLo(k_1) \; e^{2\tau_1+2\tau_2} \left[ 1 + \frac{2}{3} e^{\tau_1} \frac{\bk_1.\bw_1}{k_1^2} \right] \; \left[ 1 + \frac{2}{3} e^{\tau_2} \frac{\bk_2.\bw_2}{k_2^2} \right] .
\label{DL4}
\eeq
At lowest order, the mean $\fb$ and the propagators $G_0$ and $G$ obtained from eqs.(\ref{G0N1})-(\ref{Pi1loop1}) are:
\beq
\fb^{(0)}(x) = \frac{1}{(2\pi)^3} \; \delta_D(\bk) , \hspace{0.3cm} G_0^{(0)}(x_1,x_2) = G^{(0)}(x_1,x_2) = \DL(x_1,x_2) .
\label{foGo}
\eeq
This simply corresponds to the linear regime. On the other hand, from eqs.(\ref{R0N1})-(\ref{RN1}) we obtain for the response function $R=\lag f(\bk_1,\bw_1,\tau_1)\lambda(\bk_2,\bw_2,\tau_2)\rag_c$:
\beqa
\lefteqn{ R_0^{(0)}(x_1,x_2) = R^{(0)}(x_1,x_2) = \frac{1}{(2\pi)^6} \; \delta_D(\bk_1+\bk_2) \; \theta(\tau_1-\tau_2) \Biggl \{ \delta_D(\bw_1+\bw_2+3\bk_1(e^{-\tau_2}-e^{-\tau_1})) } \nonumber \\ & & + \int_{\tau_2}^{\tau_1} \d\tau \; \delta_D(\bw_2+3\bk_1(e^{-\tau_2}-e^{-\tau})) \left\{ \frac{2}{5} e^{\tau} \frac{\bk_1.\bw_1}{k_1^2} \left[ 2 e^{3\tau_1-3\tau} + 3 e^{-2\tau_1+2\tau} \right] + \frac{6}{5} \left[ e^{2\tau_1-2\tau} - e^{-3\tau_1+3\tau} \right] \right\} \Biggl \} .
\label{Ro1}
\eeqa
We can check that it is indeed causal and it satisfies both eqs.(\ref{R0N1}). Moreover, it also obeys both exact constraints (\ref{resp3}). In order to check the second condition (\ref{resp3}) one must pay attention to the large-scale cutoff of the gravitational interaction which multiplies the factor $6/5$ in the last term by $k_1^2/(k_1^2+k_c^2)$ (see eq.(\ref{kc1})). At next order, we obtain from eq.(\ref{fb1}) for the mean $\fb$:
\beq
\fb^{(1)}(x) = \tKs(x;y,z) G^{(0)}(y,z) = \frac{-1}{(2\pi)^3} \; \delta_D(\bk) \; \frac{2}{9} \; e^{6\tau} \int \d\bk' \; \PLo(k') \; \left( \frac{\bk'.\bw}{k'^2} \right)^2 ,
\label{fbpert1} 
\eeq
where $\tKs$ is the integrated vertex (\ref{tK1}). Note that the integral over $\bk'$ diverges at long wavelengths for $n \leq -1$. This corresponds to the infrared divergences discussed in Sect.\ref{Long-wavelength divergences} and Sect.\ref{Infrared divergences}. Next, in order to compute the next-to-leading contribution $G^{(1)}$ to the correlation $G$ we first obtain the self-energy at lowest order from eqs.(\ref{Sigma1loop1})-(\ref{Pi1loop1}). Substituting these expressions into eq.(\ref{GN1}) we can derive $G^{(1)}$. Here we are only interested in the density fluctuations hence we define:
\beq
G(\bk_1,0,\tau_1;\bk_2,0,\tau_2) = \delta_D(\bk_1+\bk_2) G(k_1;\tau_1,\tau_2) \hspace{0.3cm} \mbox{whence} \hspace{0.3cm} P(k,\tau)=(2\pi)^6 G(k;\tau,\tau) ,
\label{Gdens1}
\eeq
where $P(k,\tau)$ is the non-linear power-spectrum. Thus, $G(k;\tau_1,\tau_2)$ yields the two-point correlation of the density field (in spatial Fourier space) at times $\tau_1$ and $\tau_2$ (indeed $\bw=0$ corresponds to the integration over the momentum $\bp$). After a tedious calculation, we obtain from eq.(\ref{GN1}):
\beqa
G^{(1)}(k;\tau_1,\tau_2) & = & \theta(\tau_1-\tau) 6 (1-e^{\tau-\tau_1}) G^{(1)}(k;\tau,\tau_2) + \frac{1}{(2\pi)^6} \PLo(k) \PLo(k') e^{6\tau_1+2\tau_2} \frac{18}{7} F_3^{(s)}(\bk',-\bk',\bk) \nonumber \\ & & + \frac{1}{(2\pi)^6} \PLo(\bk-\bk') \PLo(k') e^{4\tau_1+4\tau_2} \frac{7}{5} F_2^{(s)}(\bk',\bk-\bk')^2 ,
\label{Gpert2} 
\eeqa
where the symmetric kernels $F_2^{(s)}$ and $F_3^{(s)}$ are those which appear in the standard perturbative expansion derived from the hydrodynamical approach (e.g., \cite{Jain1}). In particular, we have:
\beq
F_2^{(s)}(\bk',\bk-\bk') = \frac{5}{7} + \frac{1}{2} \frac{\bk'.(\bk-\bk')}{k'^2} + \frac{1}{2} \frac{\bk'.(\bk-\bk')}{|\bk-\bk'|^2} + \frac{2}{7} \frac{[\bk'.(\bk-\bk')]^2}{k'^2 |\bk-\bk'|^2}
\label{F2s1}
\eeq
and:
\beq
F_3^{(s)}(\bk',-\bk',\bk) = \frac{ 10 k'^4 k^6 + 10 k^4 k'^6 - 21 k^6 (\bk.\bk')^2 - 44 k^4 k'^2 (\bk.\bk')^2 - 59 k^2 k'^4 (\bk.\bk')^2 + 76 k^2 (\bk.\bk')^4 + 28 k'^2 (\bk.\bk')^4 } {126 k'^4 k^2 |\bk-\bk'|^2 |\bk+\bk'|^2 }
\label{F3s1}
\eeq
Next, the symmetric solution of the linear equation (\ref{Gpert2}) is simply:
\beq
G^{(1)}(k;\tau_1,\tau_2) = \frac{e^{6\tau_1+2\tau_2}+e^{2\tau_1+6\tau_2}}{(2\pi)^6} \PLo(k) \PLo(k') 3 F_3^{(s)}(\bk',-\bk',\bk) + \frac{e^{4\tau_1+4\tau_2}}{(2\pi)^6} \PLo(\bk-\bk') \PLo(k') 2 F_2^{(s)}(\bk',\bk-\bk')^2
\label{Gpert3} 
\eeq
Therefore, we obtain for the non-linear power-spectrum up to order $\PLo^2$:
\beq
P(k,\tau) = e^{4\tau} \PLo(k) + e^{8\tau} \left[ \PLo(k) \PLo(k') 6 F_3^{(s)}(\bk',-\bk',\bk) + \PLo(\bk-\bk') \PLo(k') 2 F_2^{(s)}(\bk',\bk-\bk')^2 \right] + \cO(\PLo^3)
\label{Ppert1}
\eeq
We can see that we recover the results of the standard perturbative approach (e.g., \cite{Gor1}, \cite{Jain1}). In particular, the integration over $\bk'$ in eq.(\ref{Ppert1}) no longer diverges at long wavelengths. Therefore, the system (\ref{G0N1})-(\ref{Pi1loop1}) is well-behaved with respect to these infrared divergences and yields meaningful results. However, note that we must still have a large-scale cutoff, or $n>-1$ on large scales, so that the intermediate quantities $\fb$ and $G(x_1,x_2)$ are finite. Besides, we explicitly see in eq.(\ref{Gpert3}) that the cancellation of the infrared divergences only holds for the equal-time correlation of the density field. This is natural, since by looking at the density field at two different times $\tau_1 \neq \tau_2$ we are obviously sensitive to the mean value of the velocity field which translates the small-scale density fluctuations over some non-zero distance during this finite time-interval.

Next, we can easily see that the same analysis can be performed at next-to-leading order for the 2PI effective action equations derived in Sect.\ref{2PI effective action}. Therefore, both sets of equations are well-behaved with respect to the infrared divergences. However, note that within the direct steepest-descent approach although the actual non-linear correlation $G$ shows the correct infrared cancellations this is not the case for the intermediate propagator $G_0$, but this is not a problem since $G_0$ has no physical meaning.

Finally, we can note that eq.(\ref{Gpert3}) confirms that solving the Vlasov equation through a perturbative expansion over powers of the linear power-spectrum $\PLo(k)$ recovers the results obtained from the hydrodynamical approach. This is consistent with the analysis presented in \cite{Val1}. This also clearly shows that we must solve exactly the equations (\ref{G0N1})-(\ref{Pi1loop1}) or (\ref{Gamfb})-(\ref{Pi2loop1}) in order to obtain interesting results. As compared with the perturbative expansion over powers of $\PLo(k)$, this automatically performs some partial infinite resummations and it is expected to be free of the ultraviolet divergences encountered in the former method. Within the 2PI effective action approach, the advantage of computing exactly the mean $\fb$ and the correlations $G$ and $R$, once we have derived $\tGam[\cG]$ up to some finite order, is that it preserves the variational principle (\ref{corr2}).

\section{Conclusion and discussion}
\label{Conclusion}

In this article we have developed a new approach to investigate the dynamics of gravitational clustering in an expanding universe. The main results of this paper are the following.

- We discussed the long-wavelength divergences which may appear for some initial conditions (Sect.\ref{Long-wavelength divergences}) and we showed how they could be treated by a simple modification of the equations of motion. However, this leads to a loss of homogeneity; hence it is more convenient to keep a large scale cutoff and to make sure that any method used in this case shows the same infrared cancellations as the standard hydrodynamical perturbative expansion.

- We derived the actions $S[f]$ and $S[f,\lambda]$ which allow one to obtain the statistical properties of the phase-space distribution function $f(\bx,\bp,t)$ from a path-integral formalism (Sect.\ref{Derivation of the actions}). We noticed that the auxiliary field $\lambda$ generates the response functions of the system.

- We briefly discussed several approximation schemes which may be applied to this functional formalism (Sect.\ref{Approximation schemes}). We recalled that a perturbative expansion over the non-Gaussian terms only recovers the standard hydrodynamical results. Then, we showed that the Feynman variational method cannot be applied to our system because of the singularities of the action $S[f]$. Next, we presented a ``semiclassical'' approach and an $N-$field generalization which may serve as a basis for large$-N$ expansions.

- We derived the equations of motion up to next-to-leading order within a direct steepest-descent method for gravitational clustering (Sect.\ref{Direct steepest-descent method}). We showed that this is possible despite the singular behaviour of the action $S[f]$.

- We presented another scheme (based on the 1PI effective action) which is expected to be more powerful than the previous steepest-descent method by analogy with other physical problems (Sect.\ref{1PI Effective action}). In particular, we compared both approaches (with some emphasis on explicit versus implicit equations) as well as the BBGKY hierarchy.

- We eventually derived for the gravitational dynamics the equations at next-to-leading order given by a 2PI effective action approach (Sect.\ref{2PI effective action}). This is the most elegant of these three methods, which we compared in Sect.\ref{Comparison with the 1PI effective action approach}.

- We showed that the infrared divergences must be taken into account and select the ``semiclassical'' approach over the $N-$field generalization (Sect.\ref{Infrared divergences}). Indeed, only the former scheme shows the right infrared cancellations. In fact, the ``semiclassical'' method agrees with the standard perturbative results up to the order to which it is derived (and it includes in addition a partial resummation of higher-order terms).

- We checked that a perturbative solution of the equations obtained from a large-$N$ approach up to next-to-leading order indeed recovers the perturbative hydrodynamical results at this same order (Sect.\ref{Perturbative expansion}). We noticed that the large-scale regularization of the gravitational interaction must be kept in mind, and that the infrared divergences which occur for some initial conditions only cancel for the equal-time density correlation.

Finally, we note that the direct steepest-descent method and the effective action approaches can be compared in the light of the virial scalings which are a key feature of gravitational systems. Thus, an isolated gravitational system obeys at equilibrium the property $2T+V=0$, where $T$ (resp. $V$) is its kinetic (resp. potential) energy. In a cosmological context, halos are not isolated nor at perfect equilibrium but they still obey a ``virial scaling'' $T \sim |V|$. Thus, after an overdensity has collapsed, the physical velocity scale $v$ obeys $v^2 \sim \cG M/R \sim \cG \rho R^2$ within an object of mass $M$, size $R$ and density $\rho$. This ``equilibrium'' (even though not perfect) between the kinetic and potential energies plays a key role in the non-linear regime of large-scale structure formation where such virialized objects build up (e.g., clusters or even filaments). Then, one can see that the implicit Schwinger-Dyson equations (\ref{Dys3}) or (\ref{varG1}) for $G$ ensure that the ``virial scaling'' will be satisfied, provided one can attach a unique time-scale and velocity-scale to a given length-scale, since the self-energy explicitly depends on the non-linear two-point correlation $G$ (of course this ``virial scaling'' is also apparent in the exact BBGKY hierarchy). Thus, at next-to-leading order, the Schwinger-Dyson eqs.(\ref{GN1})-(\ref{RN1}) with the self-energy (\ref{Sigma2loop1})-(\ref{Pi2loop1}) yield through a simple dimensional analysis:
\beq
\cO G \sim \Ks \Ks R G G \hspace{0.4cm} \mbox{whence} \hspace{0.4cm}  1 \sim \left( t_k k^3 \frac{w_k}{k} \right)^2 \; f^2 , \hspace{0.4cm} f(k,w_k) \sim t_k^{-1} k^{-2} w_k^{-1} \hspace{0.4cm} \mbox{and} \hspace{0.4cm} \rho(x) \sim k/(t_k w_k) \sim t_k^{-2} ,
\label{virial1}
\eeq
where we assumed that the time-scale $t_k$ and the velocity-scale $1/w_k$ are associated with the length-scale $x=1/k$ (whence $t_k \sim w_k/k$). The result $\rho \sim 1/t_k^2$ is the ``virial scaling'', where $\rho$ is the real space density and $t_k$ would also be called the local dynamical time. By contrast, within the direct steepest-descent scheme it is clear that in the regime where $G$ becomes very different from $\Go$ the explicit equations (\ref{Sigma1loop1})-(\ref{Pi1loop1}) imply an increasingly large deviation from the ``virial scaling''. This means that the direct steepest-descent method may not be able to handle the non-linear regime since it generically misses the ``virial scaling''. On the other hand, the effective action schemes automatically recover this behaviour (which is clearly a very important non-perturbative property of the system) because of their implicit form. Note however that this analysis does not mean that the ``virial scaling'' holds for any scale $k$. Indeed, as seen in (\ref{virial1}) in order to derive this property one needs to assume that the physics at scale $k$ is governed by this local scale and some associated time-scale $t_k$ and velocity-scale $1/w_k$. However, this is not necessarily the case because in the context of large-scale structure formation in an expanding universe there are always a characteristic time $t_H$ (the Hubble time) and a characteristic scale $R_0$ (the scale which is just turning non-linear) so that the dimensional analysis above is not sufficient (e.g., one can introduce any dependence on the dimensionless factor $kR_0$). However, we can expect the ``virial scaling'' to hold at scale $R_0$ where $t_k \sim t_H$.

Of course, the actual test of the interest of the methods developed in this paper will come from a numerical computation of the equations of motion derived here. This is in itself a non-trivial task, since the basic object is the phase-space distribution $f(\bx,\bp,t)$ which depends on seven coordinates. We plan to investigate such a numerical calculation in future works. On the other hand, one may apply further approximations to the equations of motion derived in this article at next-to-leading order over $1/N$. Finally, we note that the path-integral framework developed in this article may also serve as a basis to other tools borrowed from field theory than the ``large$-N$ methods'' we focused on.

\appendix

\section{Reduced 2PI effective action $\tGam$}
\label{Reduced 2PI effective action}

In this appendix we derive eqs.(\ref{tGam1})-(\ref{tGam2}) for a cubic action. That is, we show that the reduced effective action $\tGam$ defined in eq.(\ref{tGam1}) is indeed given by the 2PI vacuum diagrams of the action $\tS[\phi]$ defined in eq.(\ref{tGam2}). The case of a cubic action is much simpler than the general case since $\tGam[\varphi,G]$ no longer depends on $\varphi$. Thus, we present below a diagrammatic proof of eqs.(\ref{tGam1})-(\ref{tGam2}) which is much simpler than the method described in \cite{Jack1}-\cite{CJT}. We actually follow the analysis used in \cite{Vas1} to show that the full effective action $\Gamma[\varphi,G]$ only contains 2PI diagrams. Indeed, for a cubic action this method also allows one to show that the reduced effective action $\tGam[G]$ is given by a sum of 2PI vacuum diagrams. To simplify the expressions, in this appendix we take $N=1$ so that we write:
\beq
Z[j,M] = e^{W[j,M]} = \cN \int [\d\phi(x)] \; e^{j.\phi + \frac{1}{2} \phi.M.\phi - S(\phi)} \hspace{0.4cm} \mbox{with} \hspace{0.4cm} S[\phi] = S_1.\phi + \frac{1}{2} \phi.\Dm.\phi + \frac{1}{3!} S_3.\phi^3
\label{ApZ1}
\eeq
The action $S[\phi]$ is identical to eq.(\ref{Zphi1}) with non-local terms and we defined the ``bare propagator'' $\Delta$ by $\Dm=S_2$. Besides, we normalize the generating functionals $Z[j,M]$ and $Z[j]$ (the latter being defined as in eq.(\ref{Zphi1}), that is without the symmetric external matrix $M$) by the factor $\cN$ defined by:
\beq
\cN(\Delta) \int [\d\phi] \; e^{-\frac{1}{2}\phi.\Dm.\phi} = 1 , \hspace{0.4cm} \mbox{hence} \hspace{0.4cm} \cN(\Delta) = (\Det \Dm)^{1/2} .
\label{ApnormZ1}
\eeq
Thus, if $S_1=0$ and $S_3=0$ we have $Z(j=0)=Z(j=0,M=0)=1$. On the other hand, both functionals are related by:
\beq
Z(\Delta;j,M) = \frac{\cN(\Delta)}{\cN(\Db)} \; Z(\Db;j) \hspace{0.4cm} \mbox{with} \hspace{0.4cm} \Dbm = \Dm-M ,
\label{Db1}
\eeq
where we wrote explicitly the bare propagator corresponding to each generating functional $Z$ and we introduced the new propagator $\Db$. Then, the generating functionals $W=\ln Z$ of the connected correlations are related by:
\beq
W(\Delta;j,M) = W(\Db;j) + \frac{1}{2} \ln \Det (\Dm.\Db) =  W(\Db;j) - \frac{1}{2} \Tr \ln (\Delta.\Dbm) .
\label{WDb1}
\eeq
Next, using the fact that the integral of a derivative vanishes we obtain as in eqs.(\ref{Dys1})-(\ref{Dys11}):
\beq
j - S_1 + (M-\Dm).\frac{\delta W}{\delta j} - \frac{1}{2} S_3 . \left[ \frac{\delta W}{\delta j}\frac{\delta W}{\delta j} + \frac{\delta^2 W}{\delta j \delta j} \right] = 0 \hspace{0.5cm} \mbox{and} \hspace{0.5cm} \frac{\delta W}{\delta j}\frac{\delta W}{\delta j} + \frac{\delta^2 W}{\delta j \delta j} = \frac{\delta W}{\delta M} .
\label{ApDys1}
\eeq
The first equation in (\ref{ApDys1}) is simply the Schwinger-Dyson equation (\ref{Dys11}) while the second one is obtained by taking the derivatives of $Z[j,M]$ defined in eq.(\ref{ApZ1}). Indeed, since $Z[j,M]$ depends on two variables $j,M$, while the generating functional $Z[j]$ defined in eq.(\ref{Zphi1}) was a functional of the single test field $j$, we need two equations to fully define $Z[j,M]$. As pointed out in \cite{Vas1} the second eq.(\ref{ApDys1}) is actually a constraint equation which does not depend on the action $S[\phi]$. Next, we introduce the double Legendre transform $\Gamma[\varphi,G]$ as in eq.(\ref{Gam2def}) with $N=1$. The functionals $W$ and $\Gamma$ are also related by the properties:
\beq
1 = \frac{\delta\varphi}{\delta\varphi} = \frac{\delta\varphi}{\delta j} . \frac{\delta j}{\delta\varphi} + \frac{\delta\varphi}{\delta M} . \frac{\delta M}{\delta\varphi} \hspace{0.4cm} \mbox{which yields} \hspace{0.4cm} 1 = \frac{\delta^2 W}{\delta j \delta j} . \left( \frac{\delta^2 \Gamma}{\delta\varphi\delta\varphi} - 2 \frac{\delta\Gamma}{\delta G} - 2 \frac{\delta^2 \Gamma}{\delta G \delta\varphi} .\varphi \right) + \frac{\delta^2 W}{\delta j \delta M} . 2 \frac{\delta^2 \Gamma}{\delta\varphi\delta G}
\label{inv1}
\eeq
and:
\beq
0 = \frac{\delta\varphi}{\delta G} = \frac{\delta\varphi}{\delta j} . \frac{\delta j}{\delta G} + \frac{\delta\varphi}{\delta M} . \frac{\delta M}{\delta G} \hspace{0.4cm} \mbox{which yields} \hspace{0.4cm} 0 = \frac{\delta^2 W}{\delta j \delta j} . \left( \frac{\delta^2 \Gamma}{\delta\varphi\delta G} - 2 \frac{\delta^2 \Gamma}{\delta G \delta G} .\varphi \right) + \frac{\delta^2 W}{\delta j \delta M} . 2 \frac{\delta^2 \Gamma}{\delta G \delta G}
\label{inv2}
\eeq
where we used eq.(\ref{invers1}): $M=2\delta\Gamma/\delta G$ and $j=\delta\Gamma/\delta\varphi-2(\delta\Gamma/\delta G) .\varphi$. Combining both equations (\ref{inv1})-(\ref{inv2}) we obtain:
\beq
1 =  \frac{\delta^2 W}{\delta j \delta j} . \left( \frac{\delta^2 \Gamma}{\delta\varphi\delta\varphi} - 2 \frac{\delta\Gamma}{\delta G} - \frac{\delta^2 \Gamma}{\delta\varphi\delta G} . \left( \frac{\delta^2 \Gamma}{\delta G \delta G} \right)^{-1} . \frac{\delta^2 \Gamma}{\delta G \delta\varphi} \right)
\eeq
Therefore, the equations of motion (\ref{ApDys1}) also read:
\beq
\frac{\delta\Gamma}{\delta\varphi} - S_1 - \Dm.\varphi - \frac{1}{2} S_3 \left[ \varphi \varphi + G \right] = 0  \hspace{0.4cm} \mbox{and} \hspace{0.4cm} \Gm = \frac{\delta^2 \Gamma}{\delta\varphi\delta\varphi} - 2 \frac{\delta\Gamma}{\delta G} - \frac{\delta^2 \Gamma}{\delta\varphi\delta G} . \left( \frac{\delta^2 \Gamma}{\delta G \delta G} \right)^{-1} . \frac{\delta^2 \Gamma}{\delta G \delta\varphi}
\label{ApDys2}
\eeq
Of course, the first equation in (\ref{ApDys2}) is consistent with the equation of motion we had derived in eq.(\ref{Dys2}) for the first Legendre transform $\Gamma[\varphi]$, using the last property of (\ref{invers1}) to relate $\Gamma[\varphi]$ to $\Gamma[\varphi,G]$, and with eq.(\ref{varphi1}) written at the saddle-point. We can note that the first equation in (\ref{ApDys2}) is easily solved as: $\Gamma[\varphi,G] = S[\varphi] + (1/2) S_3 \varphi G - F(G)$ where $F(G)$ is an arbitrary functional of $G$. Then, $F(G)$ is determined by the second equation (\ref{ApDys2}). Therefore, we can write $\Gamma[\varphi,G]$ under the form:
\beq
\Gamma[\varphi,G] = S[\varphi] + \frac{1}{2} \Tr \left[ \Gom.G -1 \right] - \frac{1}{2} \Tr \ln \left( \Dm . G \right) - \tGam[G] \hspace{0.4cm} \mbox{with} \hspace{0.4cm} \Gom = \frac{\delta^2 S}{\delta\varphi \delta\varphi} = \Dm + S_3.\varphi 
\label{AptGam1}
\eeq
This is indeed identical to the form (\ref{tGam1}) which we want to prove here for a cubic action. Thus, we have seen so far that the reduced effective action $\tGam$ only depends on $G$ as $\Gamma[\varphi,G]$ as written in eq.(\ref{AptGam1}) is a solution of the first equation in (\ref{ApDys2}). The reason why we explicitly subtract the terms $\Tr [\Dm.G-1]$ and $\Tr\ln (\Dm.G)$ from $\tGam$ is that for a Gaussian action (i.e. $S_3=0$) we can easily solve for $\Gamma[\varphi,G]$ which yields the expression (\ref{AptGam1}) with $\tGam=0$, see also \cite{Vas1}. Then, substituting eq.(\ref{AptGam1}) into the second equation (\ref{ApDys2}) we obtain:
\beq
\frac{\delta \tGam}{\delta G} = \frac{1}{4} S_3 . \left( \Gm \Gm - 2 \frac{\delta^2 \tGam}{\delta G \delta G} \right)^{-1} . S_3
\label{rec1}
\eeq
Thus, this equation determines the reduced effective action $\tGam$, up to a term $\tGam^{(0)}$ which is independent of both $\varphi$ and $G$. We can also check that $\tGam$ only depends on $G$ and no longer on $\varphi$. This equation can be solved iteratively (see \cite{Vas1}) using:
\beq
\left( \Gm \Gm - 2 \frac{\delta^2 \tGam}{\delta G \delta G} \right)^{-1} = GG + GG \; 2 \frac{\delta^2 \tGam}{\delta G \delta G} \; G G + GG \; 2 \frac{\delta^2 \tGam}{\delta G \delta G} \; G G \; 2 \frac{\delta^2 \tGam}{\delta G \delta G} \; G G + ..
\label{series1}
\eeq
Graphically, this also reads:
\beq
\left( \Gm \Gm - 2 \frac{\delta^2 \tGam}{\delta G \delta G} \right)^{-1} = \hspace{0.1cm} \figeq{-1ex}{0.8cm}{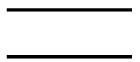} \hspace{0.1cm} + 2 \hspace{0.1cm} \figeq{-2ex}{2.35cm}{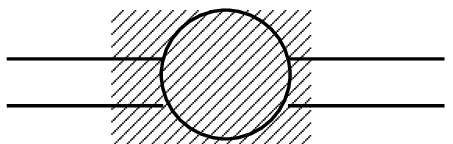} \hspace{0.1cm} + 4 \hspace{0.1cm} \figeq{-2ex}{4cm}{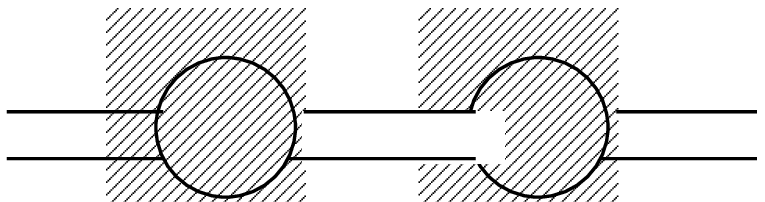} \hspace{0.1cm} + ...
\label{rec2}
\eeq
where a line stands for $G$ and a hatched circle for $\delta^2 \tGam/\delta G \delta G$. This allows us to solve for $\tGam$ as a power-series over $S_3$ (except for the zero-term $\tGam^{(0)}$ which is independent of $G$). Thus, we obtain for the first two terms:
\beq
\tGam^{(1)} =  \frac{1}{12} \hspace{0.1cm} \figeq{-2ex}{1.2cm}{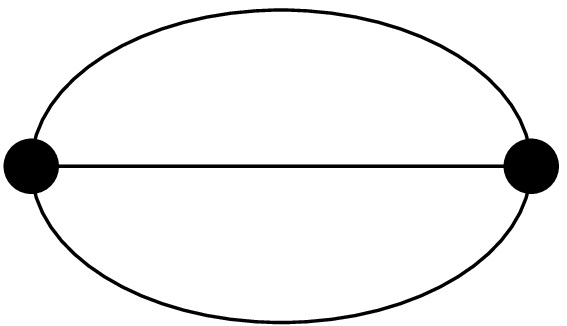} \hspace{0.1cm} , \hspace{0.4cm} \mbox{whence} \hspace{0.4cm} \frac{\delta^2 \tGam^{(1)}}{\delta G \delta G} = \hspace{0.1cm} \figeq{0ex}{1.3cm}{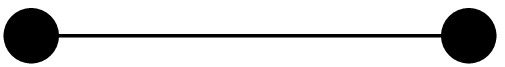} \hspace{0.4cm} \mbox{and} \hspace{0.4cm} \tGam^{(2)} =  \frac{1}{24} \hspace{0.1cm} \figeq{-2ex}{1.6cm}{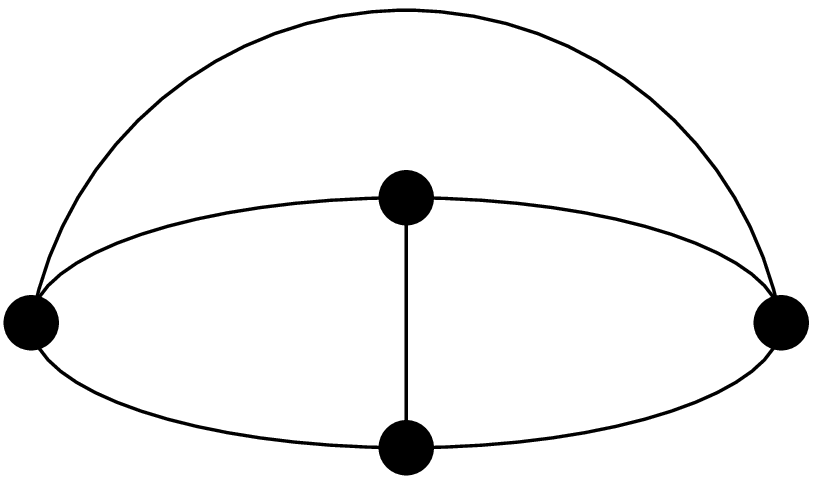} \hspace{0.2cm} .
\eeq
Here the big dot with three legs is the vertex $S_3$. We can note that these first few terms are 2PI vacuum diagrams with three-leg vertex $S_3$ and propagator $G$. Moreover, they appear with the same multiplicity factors as in $W_0(G)$, which is the term of order zero of the generating functional $W(G;j)=\sum_{n=0}^{\infty} 1/(n!) W_n j^n$ corresponding to the action $\tS[\phi] = \frac{1}{2} \phi.\Gm.\phi+\frac{1}{3!}S_3 \phi^3$. In fact, these properties remain valid at all orders. First, it is easily seen from the recursion obtained from (\ref{rec1}) and (\ref{rec2}) that all diagrams of $\tGam$ are 2PI vacuum diagrams, see \cite{Vas1} (except for the possible zero-term $\tGam^{(0)}$). Indeed, the structure of the r.h.s. in eq.(\ref{rec1}) is 2-line irreducible:
\beq
G . \frac{\delta \tGam}{\delta G} = \frac{1}{4} \; \sum_{n=0}^{\infty} \; 2^n \; \hspace{0.1cm} \figeq{-3ex}{5.5cm}{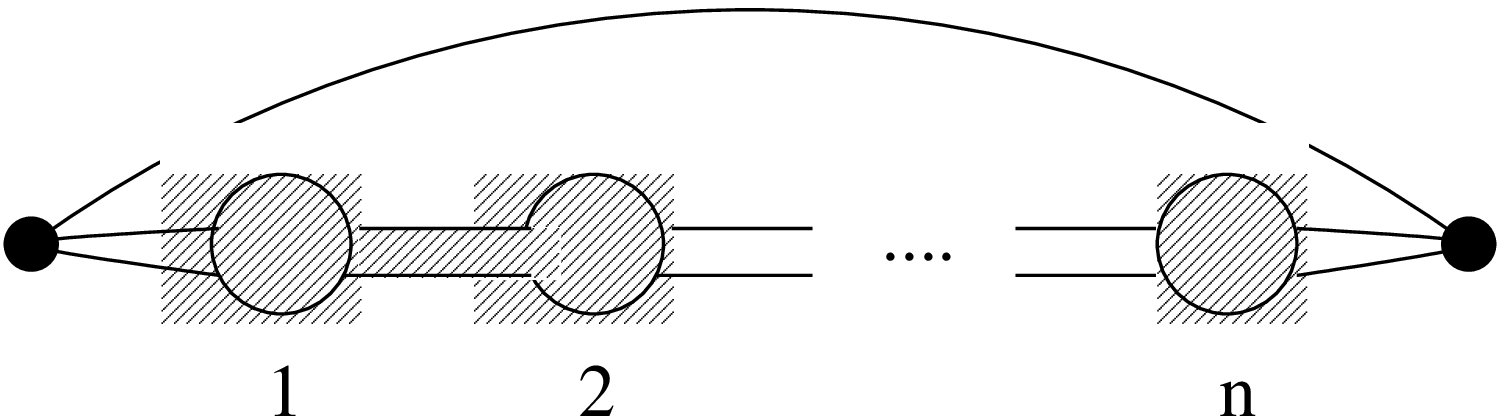}
\label{tGam2PI1}
\eeq
Second, we now need to show that $\tGam[G]$ contains all 2PI vacuum diagrams of $W_0(G)$ with the same coefficients. From the definitions of $\tGam$ and $\Gamma$ we have:
\beqa
\lefteqn{ \tGam[G] = W[j,M] - j.\varphi-\frac{1}{2} \varphi.M.\varphi -\frac{1}{2} \Tr[G.M] + S[\varphi] + \frac{1}{2} \Tr \left[ \Gom.G -1 \right] - \frac{1}{2} \Tr \ln \left( \Dm . G \right) } \nonumber \\ & & = W(\Db;j) - j.\varphi-\frac{1}{2} \varphi.M.\varphi -\frac{1}{2} \Tr[G.M] + S[\varphi] + \frac{1}{2} \Tr \left[ \Gom.G -1 \right] - \frac{1}{2} \Tr \ln \left( \Dbm . G \right)
\label{AptGam2}
\eeqa
where we used eq.(\ref{WDb1}) in the second line. Besides, $(j,M)$ are obtained from $(\varphi,G)$ through:
\beq
M= 2 \frac{\delta\Gamma}{\delta G} = \Gom - \Gm - 2 \frac{\delta\tGam}{\delta G} \hspace{0.3cm} \mbox{and} \hspace{0.3cm} j + M.\varphi = \frac{\delta\Gamma}{\delta\varphi} = S_1 + \Dm.\varphi + \frac{1}{2} S_3 [\varphi \varphi + G ] .
\label{jM1}
\eeq
Next, we can easily see from the definition of the Legendre transforms that $\Gamma_{S_1\neq 0} = \Gamma_{S_1= 0}+S_1.\varphi$ (this is obtained by noticing that the linear term $S_1.\phi$ of the action can be absorbed in the source term $j.\phi$). Since this term $S_1.\varphi$ is included in the explicit term $S[\varphi]$ in the r.h.s. in eq.(\ref{AptGam1}) we see that $\tGam[G]$ is independent of both $S_1$ and $\varphi$. Therefore, we can evaluate the r.h.s. of eq.(\ref{AptGam2}) at the point $\varphi=0, S_1=0$, which yields:
\beq
\tGam[G] = W(\Db;j) - \frac{1}{2} \Tr[G.M] + \frac{1}{2} \Tr \left[ \Dm.G -1 \right] - \frac{1}{2} \Tr \ln \left( \Dbm . G \right)
\label{AptGam3}
\eeq
with:
\beq
j = \frac{1}{2} S_3 G \hspace{0.3cm} \mbox{and} \hspace{0.3cm} M= \Dm-\Gm-2 \frac{\delta\tGam}{\delta G}, \hspace{0.3cm} \mbox{hence} \hspace{0.3cm} \Dbm= \Gm+2 \frac{\delta\tGam}{\delta G} \hspace{0.3cm} \mbox{which yields} \hspace{0.3cm} \Db = G - 2 G . \frac{\delta\tGam}{\delta G} . G + ...
\label{DbG1}
\eeq
The last equation in (\ref{DbG1}) is obtained by inverting the previous relation through an expansion over $G$. Substituting $M$ and $\Dbm$ into eq.(\ref{AptGam3}) we have:
\beqa
\tGam[G] & = & W(\Db;j) + \Tr \left[ G . \frac{\delta\tGam}{\delta G} \right] - \frac{1}{2} \Tr \ln \left( 1 + 2 G . \frac{\delta\tGam}{\delta G} \right) \nonumber \\ & = & W(\Db;j) + \Tr \left[ G . \frac{\delta\tGam}{\delta G} \right] - \frac{1}{2} \Tr \left[ 2 G . \frac{\delta\tGam}{\delta G} - 2 G . \frac{\delta\tGam}{\delta G} . G . \frac{\delta\tGam}{\delta G} + ... \right] ,
\label{AptGam4}
\eeqa
where we developed the logarithm. First, we see from eqs.(\ref{DbG1})-(\ref{AptGam4}) that $\tGam[G=0]=0$ therefore the zero-term vanishes: $\tGam^{(0)}=0$. Hence $\tGam$ only contains 2PI diagrams over $G$, as shown above from eqs.(\ref{rec1})-(\ref{tGam2PI1}), so that $\tGam = \left. \tGam \right|_{\rm 2PI}$. Second, as noticed in \cite{Vas1}, we can see that the first term of the series in eq.(\ref{AptGam4}) cancels out with the second term in the r.h.s. of eq.(\ref{AptGam4}) while higher-order terms are 2-particle reducible:
\beq
\Tr \ln \left( 1 + 2 G . \frac{\delta\tGam}{\delta G} \right) = \; 2 \hspace{0.2cm} \figeq{-3ex}{0.8cm}{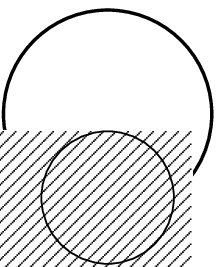} \hspace{0.2cm} - 2 \hspace{0.2cm} \figeq{-4ex}{1.3cm}{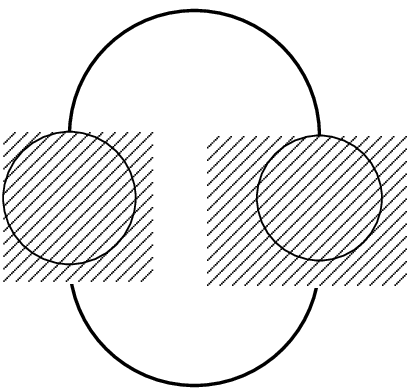} \hspace{0.2cm} + \frac{8}{3} \hspace{0.2cm} \figeq{-4ex}{1.4cm}{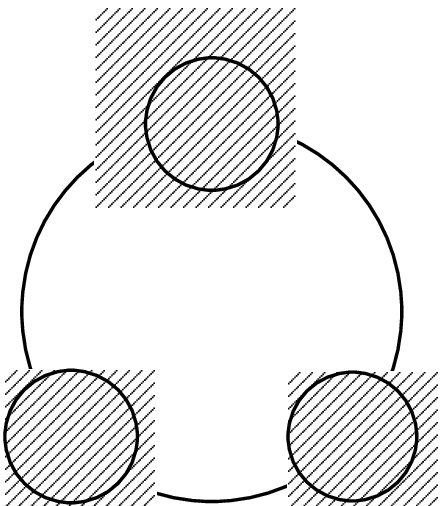} \hspace{0.2cm} + ...
\eeq
Here a line stands for $G$ and a hatched circle for $\delta\tGam/\delta G$. Therefore, we obtain:
\beq
\tGam[G] = \left. \tGam[G] \right|_{\rm 2PI} = \left. W\left(\Db;j=\frac{1}{2} S_3 G\right) \right|_{\rm 2PI} = \left. \sum_{n=0}^{\infty} \frac{1}{n!} \; W_n(\Db) . \left( \frac{1}{2} S_3 G \right)^n \right|_{\rm 2PI} ,
\label{tGamW1}
\eeq
where $\left. W \right|_{\rm 2PI}$ is the 2PI part of $W$ (expressed in terms of $G$). In eq.(\ref{tGamW1}) we used the observation that if two functionals are equal their 2-particle irreducible parts (2PI) are also equal (\cite{Vas1}). Next, we note that for $n \geq 1$ the r.h.s. in eq.(\ref{tGamW1}) contains at least one external leg attached to $j$ which implies that the diagram is 1-particle reducible:
\beq
n \geq 1 : \hspace{0.2cm} W_n . j^n \hspace{0.2cm} \Rightarrow \hspace{0.2cm} \figeq{-1.5ex}{1.1cm}{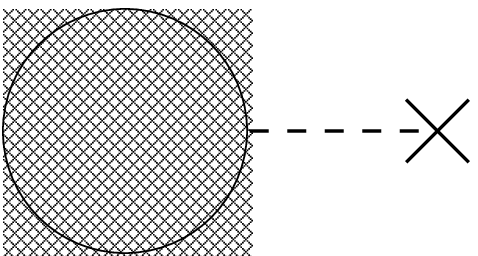} \hspace{0.1cm} = \frac{1}{2} \hspace{0.1cm} \figeq{-1.5ex}{1.2cm}{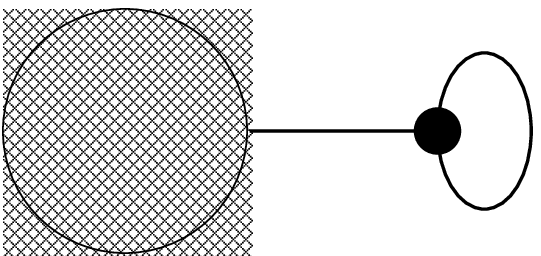} \hspace{0.1cm} - \hspace{0.1cm} \figeq{-1.5ex}{1.7cm}{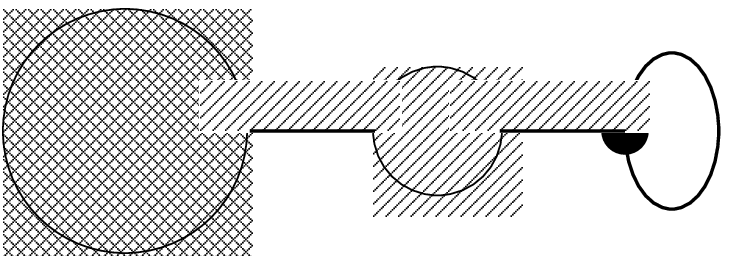} \hspace{0.1cm} + .. \hspace{0.3cm} \mbox{since} \hspace{0.3cm} j=\frac{1}{2} S_3 G \hspace{0.3cm} \mbox{and} \hspace{0.3cm} \Db = G - 2 G . \frac{\delta\tGam}{\delta G} . G + ...
\label{legj1}
\eeq
Here we singled out one branch attached to a weight $j$ (shown by the cross). The big dot with three legs is again the vertex $S_3$, the dashed line is the propagator $\Db$ and the solid line is the propagator $G$. The large filled circle on the left is $W_n.j^{n-1}$ while the middle hatched circle in the right diagram is $\delta\tGam/\delta G$. Therefore, only the first term $W_0(\Db)$ in the series in eq.(\ref{tGamW1}) (i.e. vacuum diagrams) contributes to the 2PI part of $W$. Next, we note that within these diagrams we may replace $\Db$ by $G$ since any insertion of a factor $\delta\tGam/\delta G$ yields a 2-particle reducible graph:
\beq
\left. \figeq{-1.5ex}{1.1cm}{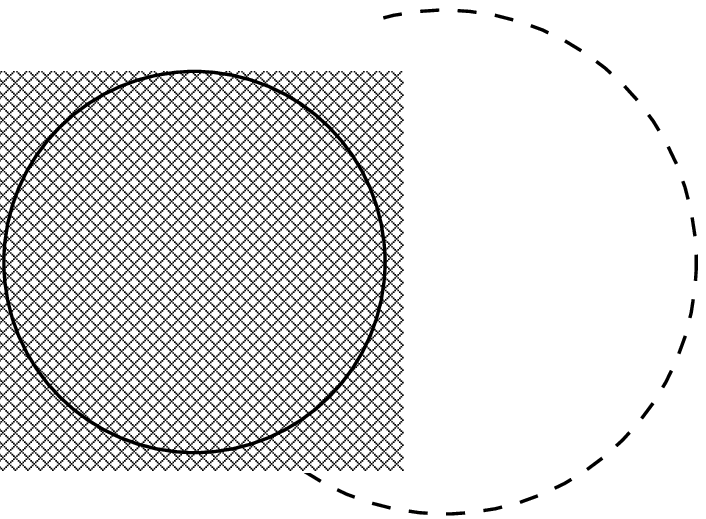} \hspace{0.1cm} \right|_{\rm 2PI} \hspace{0.1cm} = \hspace{0.1cm} \left. \figeq{-1.5ex}{1.1cm}{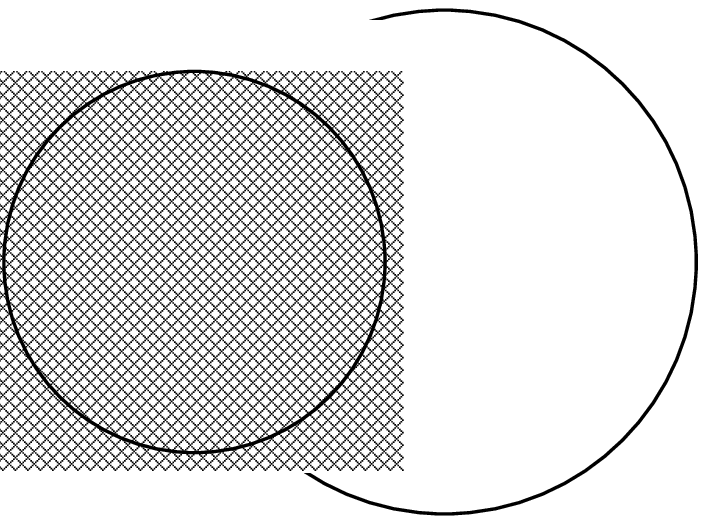} \hspace{0.1cm} - 2 \hspace{0.1cm} \figeq{-1.5ex}{1.3cm}{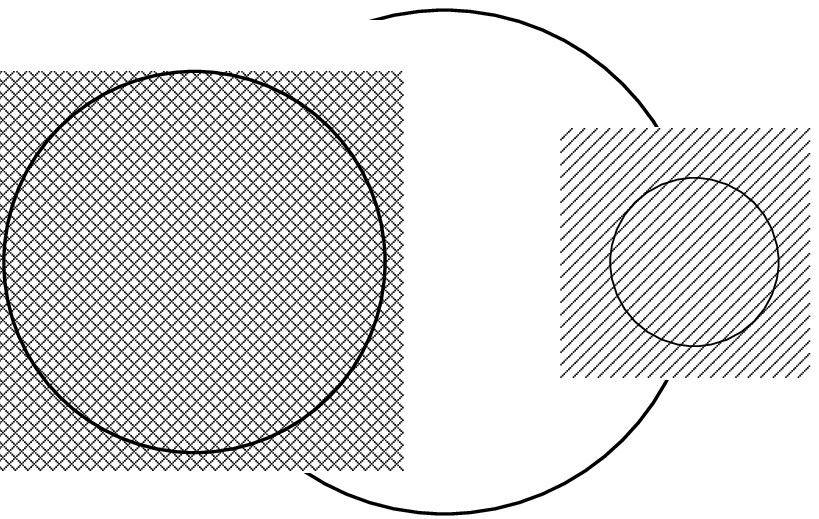} \hspace{0.1cm} + ... \right|_{\rm 2PI} \hspace{0.1cm} = \hspace{0.1cm} \left. \figeq{-1.5ex}{1.1cm}{ex2P2} \hspace{0.1cm} \right|_{\rm 2PI} .
\eeq
Therefore, we have shown (remember that $W$ must be computed with $S_1=0$ in eq.(\ref{AptGam3}) and afterwards):
\beq
\tGam[G] = \left. W_0(G) \right|_{\rm 2PI} \hspace{0.1cm} = \hspace{0.1cm} \mbox{2PI vacuum diagrams of the action} \hspace{0.1cm} \tS[\phi] = \frac{1}{2} \phi.\Gm.\phi + \frac{1}{3!} S_3 \phi^3 .
\label{tGamW01}
\eeq	
This completes the proof that for a cubic action $S[\phi]$ as given in eq.(\ref{ApZ1}), the reduced effective action $\tGam$ introduced in eq.(\ref{AptGam1}) is given by the 2PI vacuum diagrams of the action $\tS$ written in eq.(\ref{tGamW01}), with the same multiplicity factors as in $W_0$.

For the case where we factorize an integer $N$ in front of the action, as in (\ref{Zphi1}), we simply need to make the following substitutions within the formulae derived in this appendix: $S\rightarrow NS, W\rightarrow NW, j\rightarrow Nj, M\rightarrow NM, \Dm\rightarrow N\Dm, \Gamma \rightarrow N\Gamma, \varphi\rightarrow\varphi, G\rightarrow G/N$ and we obtain eqs.(\ref{tGam1})-(\ref{tGam2}).


\begin{thebibliography}{}

\bibitem[Aarts et al. (2002)]{Aarts1}
Aarts G., Ahrensmeier D., Baier R., Berges J., Serreau J., 2002,
Phys. Rev. D, 66, 45008

\bibitem[Baym (1962)]{Baym1}
Baym G., 1962, Phys. Rev., 127, 1391 

\bibitem[Berges (2002)]{Berg1}
Berges J., 2002, Nucl. Phys. A, 699, 847

\bibitem[Bernardeau et al. (1994)]{Ber2}
Bernardeau F., 1994, A\&A, 291, 697

\bibitem[Bernardeau et al. (2002)]{Ber1}
Bernardeau F., Colombi S., Gaztanaga E., Scoccimarro R., 2002, Phys. Rep., 367, 1

\bibitem[Buchert et al. (1999)]{Buch1}
Buchert T., Dominguez A., Perez-Mercader J., 1999, 349, 343

\bibitem[Calzetta \& Hu (1994)]{Cal1}
Calzetta E., Hu B.L., 1994, Heat kernel techniques and quantum gravity, Proc., Winnepeg, Canada, eds. S.A. Fulling, hep-th/9501040

\bibitem[Coles et al. (1993)]{Coles1}
Coles P., Melott A.L., Shandarin S.F., 1993, MNRAS, 260, 765

\bibitem[Coles \& Spencer (2003)]{Coles2}
Coles P., Spencer K., 2003, MNRAS, 342, 176

\bibitem[Cooper et al. (2003)]{Coop1}
Cooper F., Dawson J.F., Mihaila B., 2003, Phys. Rev. D, 67, 51901

\bibitem[Cornwall et al. (1974)]{CJT}
Cornwall J.M., Jackiw R., Tomboulis E., 1974, Phys. Rev. D, 10, 2428

\bibitem[Davies \& Widrow (1997)]{Dav1}
Davies G., Widrow L.M., 1997, ApJ, 485, 484

\bibitem[De Dominicis (1962)]{Cyr1}
De Dominicis C., 1962, J. Math. Phys., 3, 983

\bibitem[De Dominicis \& Martin (1964)]{Cyr2}
De Dominicis C., Martin P.C., 1964, J. Math. Phys., 5, 14, 31

\bibitem[Feynman (1972)]{Fey1}
Feynman R.P., 1972, Statistical mechanics, Addison Wesley Longman

\bibitem[Fry (1984)]{Fry1}
Fry J.N., 1984, ApJ, 279, 499

\bibitem[Goroff et al. (1986)]{Gor1}
Goroff M.H., Grinstein B., Rey S.-J., Wise M.B., 1986, ApJ, 311, 6

\bibitem[Gurbatov et al. (1989)]{Gur1}
Gurbatov S.N., Saichev A.I., Shandarin S.F., 1989, MNRAS, 236, 385

\bibitem[Itzykson \& Zuber (1980)]{Itz1}
Itzykson C., Zuber J.-B., 1980, Quantum field theory, McGraw-Hill

\bibitem[Jackiw (1974)]{Jack1}
Jackiw R., 1974, Phys. Rev. D, 9, 1686

\bibitem[Jain \& Bertschinger (1996)]{Jain1}
Jain B., Bertschinger E., 1996, ApJ, 456, 43

\bibitem[Jensen (1981)]{Jens}
Jensen R.V., 1981, J. Stat. Phys., 25, 183

\bibitem[Martin et al. (1973)]{MSR}
Martin P.C., Siggia E.D., Rose H.A., 1973, Phys. Rev. A, 8, 423

\bibitem[Mihaila et al. (2001)]{Mih1}
Mihaila B., Dawson J.F., Cooper F., 2001, Phys. Rev. D, 63, 096003

\bibitem[Peebles (1980)]{Peeb1}
Peebles P.J.E., 1980, The large scale structure of the universe,
Princeton University Press

\bibitem[Peebles (1982)]{Peeb2}
Peebles P.J.E., 1982, ApJ, 263, L1

\bibitem[Phythian (1977)]{Phy}
Phythian R., 1977, J. Phys. A, 10, 777

\bibitem[Scoccimarro \& Frieman (1996)]{Scoc1}
Scoccimarro R., Frieman J.A., 1996, ApJ, 473, 620

\bibitem[Valageas (2001)]{Val1}
Valageas P., 2001, A\&A, 379, 8

\bibitem[Valageas (2002a)]{Val2}
Valageas P., 2002, A\&A, 382, 412

\bibitem[Valageas (2002b)]{Val4}
Valageas P., 2002, A\&A, 382, 450

\bibitem[Valageas (2002c)]{Val5}
Valageas P., 2002, A\&A, 382, 477

\bibitem[Vasil'ev \& Kazanskii (1972)]{Vas1}
Vasil'ev A.N., Kazanskii A.K., 1972, Teor. Mat. Fiz., 12, 352

\bibitem[Vergassola et al. (1994)]{Ver1}
Vergassola M., Dubrulle B., Frisch U., Noullez A., 1994, 289, 325

\bibitem[Vishniac (1983)]{Vish1}
Vishniac E.T., 1983, MNRAS, 203, 345

\bibitem[Widrow \& Kaiser (1993)]{Wid1}
Widrow L. M., Kaiser N., 1993, ApJ, 416, L71

\bibitem[Zel'dovich (1970)]{Zel}
Zel'dovich Y.B., 1970, A\&A, 5, 84

\bibitem[Zinn-Justin (1989)]{Zinn1}
Zinn-Justin J., 1989, Quantum field theory and critical phenomena, Clarendon Press, Oxford 

\end{thebibliography}
\end{document}